\documentclass{article}

\usepackage{arxiv}

\usepackage[utf8]{inputenc} 
\usepackage[T1]{fontenc}    
\usepackage{hyperref}       
\usepackage{url}            
\usepackage{booktabs}       
\usepackage{amsfonts}       
\usepackage{nicefrac}       
\usepackage{microtype}      
\usepackage{lipsum}		
\usepackage{graphicx}
\usepackage{natbib}
\usepackage{doi}
\usepackage{amsmath,amssymb,amsfonts}%
\usepackage[title]{appendix}%

\usepackage{comment}
\usepackage{ulem}
\usepackage{bm}

\newcommand{\vect}[1]{{\bm {#1} }}
\newcommand{\ddt}[0]{\frac{d}{dt}}

\newcommand{\bra}[1]{\langle#1\rvert} 
\newcommand{\ket}[1]{\lvert#1\rangle} 
\newcommand{\qprod}[2]{ \langle #1 | #2 \rangle} 
\newcommand{\braopket}[3]{\langle #1 | #2 | #3\rangle} 

\newcommand{\op}[1]{\mathbf{ #1 }} 
\newcommand{\conj}[1]{ {#1}^\ast }

\newcommand{\wmode}{\vect{\Psi}^\pm_\textrm{w}}
\newcommand{\iomode}{\vect{\Psi}_\textrm{io}}
\newcommand{\gmode}{{\vect{\Psi}}_\textrm{g}}
\newcommand{\gOmode}{{\vect{\Psi}}_{\textrm{g}0}}
\newcommand{\mdamode}{{\vect{\Psi}}_{\textrm{mda}}}

\newcommand{\etat}{\eta}
\newcommand{\etae}{\eta_\textrm{e}}

\title{Available potential vorticity and the wave-vortex decomposition for arbitrary stratification}

\date{March 28, 2024}	

\author{ \AnotherAuthor{\href{https://orcid.org/0000-0003-4332-4569}{\includegraphics[scale=0.06]{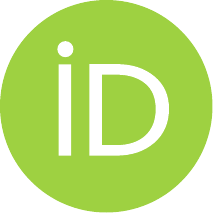}\hspace{1mm}Jeffrey J. Early}\\
	NorthWest Research Associates\\
	Seattle, WA, USA \\
	\texttt{jeffrey@jeffreyearly.com} }
  &
	\AnotherAuthor{\href{https://orcid.org/0000-0002-4845-6723}{\includegraphics[scale=0.06]{orcid.pdf}\hspace{1mm}Gerardo Hernández-Dueñas} \\
	National Autonomous University of Mexico \\
	Querétaro, Mexico \\
	\texttt{hernandez@im.unam.mx} }
 \\ & \\[1cm]
	\AnotherAuthor{\href{https://orcid.org/0000-0002-7010-5460}{\includegraphics[scale=0.06]{orcid.pdf}\hspace{1mm}Leslie M. Smith} \\
	University of Wisconsin - Madison\\
	Madison, WI, USA \\
	\texttt{lsmith@math.wisc.edu} }
 &
	\AnotherAuthor{\href{https://orcid.org/0000-0002-2562-9628}{\includegraphics[scale=0.06]{orcid.pdf}\hspace{1mm}M.-Pascale Lelong} \\
	NorthWest Research Associates\\
	Seattle, WA, USA \\
	\texttt{pascale@nwra.com} }
}

\renewcommand{\shorttitle}{APV and the wave-vortex decomposition}

\hypersetup{
pdftitle={Available potential vorticity and the wave-vortex decomposition for arbitrary stratification},
pdfsubject={q-bio.NC, q-bio.QM},
pdfauthor={Jeffrey J. Early,},
pdfkeywords={internal gravity waves, geostrophic motions, energy orthogonality, mesoscale eddies},
}

\begin{document}
\maketitle

\begin{abstract}
We consider a rotating non-hydrostatic flow with arbitrary stratification and argue that 1) the appropriate form of potential vorticity (PV) for this system is formulated in terms of isopycnal deviation and 2) the decomposition into energetically orthogonal solutions is fundamentally a PV-inversion.

The new closed-form expression for available potential vorticity (APV) is expressed in terms of isopycnal deviation, following the ideas articulated in Wagner \& Young (2015). This form of APV directly linearizes to quasigeostrophic PV after discarding the nonlinear stretching term and a height nonlinearity, the latter of which is not present in constant stratification. This formulation leads to positive definite definitions of both potential enstrophy and total energy expressed in terms of isopycnal deviation, from which the quadratic versions emerge at lowest order. It is exactly these quantities diagonalized by the linear eigenmodes.

Internal-gravity waves, geostrophic motions, inertial oscillations, and a mean density anomaly form the energetically and enstrophically orthogonal constituents of flow. The complete state of the fluid can be represented in terms of these physically realizeable modes, and determined from the derived projection operators using the horizontal velocity and density anomaly. The projection of the fluid state onto the non-hydrostatic wave modes in particular, reveals that one must first account for the PV portion of the flow, before recovering the wave solutions.

We apply the physical insights of the decomposition to a geostrophic streamfunction for a mesoscale eddy and show how strict adherence to adiabatic rearrangement of potential density places strong constraints on the vertical structure of such eddies, including an expected skew towards stronger cyclonic eddies in the upper water-column. Finally, the expression for APV is shown to reproduce the height nonlinearity of shallow-water potential vorticity, a well know feature that breaks the cyclone-anticyclone symmetry in quasigeostrophic potential vorticity.
\end{abstract}


%
\section{Introduction}
\label{sec:introduction}
%

Geostrophic motions and internal gravity waves (IGWs) form two of the primary building blocks of geophysical fluid dynamics and methods for disentangling these types of motions are often referred to as \emph{wave-vortex decompositions}.\footnote{The word `vortex' somewhat vaguely refers to the geostrophic, or balanced, part of the flow with dynamics linked to potential vorticity conservation.} Broadly speaking there are two classes of wave-vortex decompositions: full-knowledge decompositions and sparse-data decompositions. The full-knowledge decompositions assume complete knowledge of the fluid state and are then challenged with separating the fluid state into energetically orthogonal constituent parts. The challenges associated with such decompositions are  theoretical in nature, but result in separations with no ambiguities beyond methodological assumptions. Sparse-data decompositions, on the other hand, are faced with the same task of separating the fluid state into constituent parts, but with the additional constraints imposed by incomplete knowledge of the fluid state due to sparse sampling, e.g., velocity time series from a mooring. The challenges here are far more practical in nature, requiring statistical assumptions to proceed, and thus resulting in uncertainties related to the sparse sampling. 

Full-knowledge decompositions are a prerequisite to sparse-data decompositions. After all, if we cannot agree on what constitutes a wave or vortex with complete knowledge of the fluid state, how can we agree on what we are measuring from sparse observations? Thus at the heart of every wave-vortex decomposition is the idea of observability. Each constituent in the decomposition must be a measurable, meaningfully distinct fluid state. Constituents of the wave-vortex decomposition certainly must satisfy all boundary conditions and solve the equations-of-motion, but what makes a solution meaningfully distinct from other solutions? The answer in geophysical fluid dynamics is the same as in quantum mechanics: energy orthogonality.

\subsection{What is energy orthogonality?}

Energy (and enstrophy) orthogonality is an idea that underpins many of the fundamental concepts in geophysical fluids and without it, the inverse energy cascade and forward enstrophy cascade of two-dimensional turbulence would not be well defined concepts, nor would it be possible to construct an internal wave energy spectrum. The importance of energy and enstrophy orthogonality is rarely discussed explicitly, likely because in the simplest examples it follows directly from Parseval's theorem for Fourier modes. To understand the importance of this concept we spend some time drawing the analogy with Parseval's theorem for Fourier modes and discussing several examples where orthogonality has been important in geophysical fluids.

The state of a geophysical fluid (for our purposes) is given by $(u,v,w,\eta,p)$ for fluid velocity $(u,v,w)$ with isopycnal deviation $\eta$ and pressure $p$---and any basis capable of representing all physically realizeable states of the fluid is said to be \emph{complete}. Observations of these dynamical variables in space and time are rather trivially recorded on a complete basis, e.g., $n$ observations are recorded in a vector of length $n$. Representing numerical or analytical solutions with, e.g., a cosine basis, requires a bit more care, as the basis must be compatible with the dynamically imposed boundary conditions. That said, there are numerous suitable complete bases capable of representing the physically realizeable states of the system. The reverse is also true: complete bases such as cosine and Chebyshev series are suitable for a variety of different physical systems. In contrast to a complete basis, an energetically orthogonal basis is deeply tied to the dynamics.




To lowest order the total volume-integrated \emph{available energy} of the rotating, non-hydrostatic Boussinesq is
\begin{equation}
\label{eqn:intro-total-energy}
    \mathcal{E}\left[ (u,v,w,\eta) \right] = \frac{1}{2} \int \left( u^2 + v^2 + w^2 + N^2 \eta^2 \right) dV
\end{equation}
where $N^2(z)$ is the squared buoyancy frequency. If the fluid can be linearly decomposed into two parts such that $\left(u,v,w,\eta \right)=\left(u_1+u_2,v_1+v_2,w_1+w_2,\eta_1+\eta_2 \right)$ then
\begin{equation}
\label{eqn:intro-energy-sum}
    \mathcal{E}\left[\left(u,v,w,\eta \right) \right] = \mathcal{E}\left[ \left(u_1,v_1,w_1,\eta_1 \right) \right] + \mathcal{E}\left[ \left(u_2,v_2,w_2,\eta_2 \right) \right] + \epsilon_{12}
\end{equation}
where $\epsilon_{12}$ is a quadratic cross-term. For a sparse-data decomposition it may be that the best one can hope for is that the cross-term $\epsilon_{12}$ is small, but for a full-knowledge decomposition it may be possible to find solutions where the cross-term vanishes, $\epsilon_{12} = \delta_{12}$ and decomposition is \emph{energetically orthogonal}.

When a basis is both complete and energetically orthogonal then each basis member has a unique energy signature---a feature that allows one to say that energy is moving \emph{from} somewhere \emph{to} somewhere else. Moreover, the lack of cross-terms also allows us to write an energy spectrum because the sum of the squares of the basis functions is proportional to the total energy. This is exactly analogous to a Fourier series which partitions the total variance of a periodic function in terms of orthogonal Fourier modes.

The potential enstrophy of fluid is another quadratic quantity that characterizes the fluid,
\begin{equation}
    \mathcal{Z}\left[ (u,v,w,\eta) \right] = \frac{1}{2} \int \left( v_x - u_y -f \eta  \right)^2 dV
\end{equation}
and a wave-vortex decomposition will require enstrophy orthogonality following rules analogous to \eqref{eqn:intro-energy-sum}.




For two-dimensional doubly-periodic and three-dimensional triply-periodic flows all components of the energetically and enstrophically orthogonal modes are proportional to Fourier modes, and thus energy orthogonality is simply a rescaled version of Fourier mode orthogonality. Due to this almost trivial relationship between the two, it would be easy to miss the significance of energy orthogonality.

\subsection{Why are energy and enstrophy orthogonality important?}

One of the most important results in rotating stratified fluids is the existence of the inertial cascades \citep{vallis2006-book}. The central idea is that energy moves from smaller scales to larger scales while simultaneously enstrophy moves from larger scales to smaller scales. To make such claims and simultaneously observe this dual cascade requires a basis of eigenmodes that are both energetically and enstrophically orthogonal.

The formulation of the oceanic internal wave spectrum also hinges on energy orthogonality. The canonical form of the internal wave spectrum, the Garrett-Munk (GM) spectrum, is constructed in a basis of energetically orthogonal hydrostatic internal wave eigenmodes \citep{garrett1972-gfd}, such that the depth integrated total energy is expressed as a sum of squares of individual eigenmodes. Importantly, energy orthogonality is not just a choice, but a requirement for constructing an energy spectrum. Approximating the hydrostatic eigenmodes with the WKB approximation leads to a formulation of the GM in terms of a stretched coordinate where the eigenmodes look like rescaled Fourier modes and energy orthogonality resembles Fourier mode orthogonality again.

While the formulation of the internal gravity wave spectrum by \citet{garrett1972-gfd} is sufficient for describing the energy content of interior flow, eigenmodes with an explicit free-surface are necessary when surface (or barotropic) waves are present. \citet{kelly2016-jpo} constructed an energetically orthogonal basis for hydrostatic waves with an explicit free-surface using vertical modes first described by \citet{olbers1986-igw}. This solved a decades long `spurious energy conversion' problem which, when viewed through the lens of energy orthogonality, was caused by attempts to use a non-orthogonal basis.

To assess the energy and enstrophy fluxes in three dimensions with variable stratification and buoyancy anomalies at the boundaries, requires an energetically enstrophically orthogonal basis. Finding a such a basis is not trivial. The basis proposed by \citet{smith2013-jpo} satisfies these conditions for quasigeostrophic flows.

With an energetically orthogonal basis at hand, it then becomes possible to identify the physically important mechanisms that shape the observed spectra and precisely quantity the role of each mode in the transfer of energy. These include methods that limit the available interactions \citep{hernandez2014-jfm,eden2019-jpo}, directly diagnose energy transfers \citep{eden2020-jpo}, or even directly attribute the mechanisms responsible for stirring \citep{hernandez2021-jpo}.

Methodologies for separating dynamical effects that formally lack energy orthogonality may very well be connected to an energy-orthogonal decomposition. \citet{lelong2020-jpo} formulate a hybrid spatial-temporal decomposition to isolate the energetics of a mesoscale eddy, inertial oscillations, and a developing internal gravity wave field. The energetic transfers were consistent with the hypothesis that the eddy acted at a catalyst in facilitating the transfer of energy from the inertial oscillations to the internal gravity wave field. With a complete energetically orthogonal basis, \citet{early2021-jfm} measured this and the other fluxes exactly, confirming the hypothesis directly. Other methodologies for isolating dynamical phenomena based on filtering or scaling are almost certainly related to full-knowledge wave-vortex decompositions.

\subsection{A march towards realism}

The prototypical wave-vortex decompositions are either triply-periodic with constant stratification or shallow-water models---both of which are highly idealized. Although the single-layer shallow-water model would appear to be the simpler one, its energy and enstrophy both have non-quadratric terms due to the height nonlinearity \citep{remmel2009-jfm}, a feature which is also present in the variable stratification model presented in this manuscript. Overall, the community continues to pursue a number of different directions that move beyond these prototypical models and incorporate increasingly physically realistic effects.

A complete, energetically orthogonal solution set unambiguously disentangles the fluid state into constituent wave and geostrophic motions, but the lack of ambiguity is only as good as the physical interpretation of the linear eigenmodes. One line of inquiry is to extend the linear eigenmode decomposition to incorporate higher order effects.  It has long been recognized that true balanced motions require nonlinear corrections proportional to the Rossby number and thus a true separation requires additional steps beyond normal mode decomposition \citep{chouksey2023-jfm}.

At larger scales order Rossby number effects remain small. Recent work has extended the normal mode decomposition to the shallow-water equations on the sphere \citep{vasy2021-qjrms}. The geostrophic mode is now interpreted as a Rossby mode, and the methodology has been applied both numerically and diagnostically.

Another glaring omission of the prototypical wave-vortex decompositions is the lack of variable stratification. Recent work has extended the wave-vortex decompositions to include non-hydrostatics with variable stratification \citep{early2021-jfm}. Just like shallow-water, energy and enstrophy now contain higher order effects not present in constant stratification.

Numerous other physical effects remain to be incorporated into wave-vortex decompositions. \citet{oneill2024-jtech} develop a methodology that includes Ekman effects at the ocean surface boundary. Although the methodology lacks formal proof of orthogonality, it successfully diagnoses fluid flow at the boundary. This and other effects will need to be incorporated into wave-vortex decompositions to make more direct contact with sparse-data decompositions. 

\subsection{The work here}

The work presented here continues the march towards increasing realism in wave-vortex decompositions by resolving a number of outstanding issues surrounding the variable stratification decomposition, including the correct definition of available potential vorticity and orthogonality of non-hydrostatic eigenmodes. We focus particularly on observability, ensuring that all states of motion are physically realizeable. In practice, this means that each eigen-state satisfies the linear equations of motion, all boundary conditions, and even all quadratic conservation laws. A notable exception to this is the mean-density anomaly (mda) mode, discussed at length below.

The wave-vortex decompositions hinge on energy and enstrophy orthogonality of the linear solutions and thus expressions for the fully nonlinear energy and potential enstrophy are critical for articulating the order of approximation in the linear solutions. As noted in \citep{early2021-jfm} however, expressions for potential vorticity either contain a large background contribution and thus do not linearize to the correct quantity (i.e., Ertel PV), or do not have closed form solutions \citep{wagner2015-jfm}. Here we present what we claim is the `best' definition for available potential vorticity (APV) in variable stratification on an $f$-plane, which matches the intention of \citet{wagner2015-jfm}, but in closed form. This expression first appeared in \citep{muller1995-rg}, although with erroneous caveats limiting its applicability as we noted in \citep{early2022-arxiv}. The key to our expression for APV is the use of isopycnal deviation as the dependent variable, just as in \citep{holliday1981-jfm} for the construction of available potential energy (APE). The nonlinear conservation laws for APV and APE include nonlinearities that are more akin to their shallow-water versions, rather than three-dimensional constant stratification. The conservation laws including their volume integrated forms, are described in section \ref{sec:mat-cons-of-z-eta}.

Although the nonlinear conservation laws are most naturally expressed in terms of isopycnal deviation, the linearized conservation laws and equations-of-motion are best formulated in terms in terms of excess density, as described in section \ref{sec:linearization}. The complete energetically and enstrophically orthogonal solution set is presented in section \ref{sec:orthogonal-solutions}. The solution set includes two types of waves modes devoid of potential vorticity, inertial oscillations and internal gravity wave modes, and two types of potential vorticity carrying modes, the mean-density anomaly (mda) mode and the geostrophic modes. The mean-density anomaly modes  describe the difference between the mean density density state and the flattened isopycnal states of motion. These modes carry both potential enstrophy and potential energy and are necessary for a complete description of the fluid state.

With a complete energetically orthogonal basis at hand, section \ref{sec:nonlinear-wave-vortex} expresses the nonlinear equations of motion in terms of this wave-vortex basis. Reduced interaction models follow immediately, including stratified quasigeostrophic potential vorticity equation by restriction to the potential vorticity carrying modes, and wave-only models by restriction to the wave modes. 

Finally section \ref{sec:applications} shows the implications of these results on our understanding of mesoscale eddies. In particular, strict adherence to the boundary conditions implies that  nonlinear mesoscale eddies will favor cyclones, especially with density anomalies near the surface, consistent with past results. The new form of APV includes a height nonlinearity not present in constant stratification that resembles the height nonlinearity in shallow-water. This nonlinear correction is of $O(1)$ importance, as has long been predicted for geostrophic motions at the mesoscales. Section \ref{sec:discussion} provides discussion and conclusions.

%
\section{Background}
\label{sec:background}
%

Under the Boussinesq approximation, we consider the inviscid equations of motion
\begin{subequations}
\label{eqn:boussinesq}
\begin{align}
\label{x-momentum}
 \frac{du}{dt} - f v   =& - \frac{1}{\rho_0} \partial_x p_\textrm{tot}\\ \label{y-momentum}
 \frac{dv}{dt} + f u  =&  - \frac{1}{\rho_0}\partial_y p_\textrm{tot}  \\ \label{z-momentum}
 \frac{dw}{dt}  =& - \frac{1}{\rho_0}\partial_z p_\textrm{tot} -\frac{1}{\rho_0}g \rho_\textrm{tot} \\ \label{thermodynamic}
\frac{d \rho_\textrm{tot}}{dt}=& 0  \\ 
\label{continuity}
\partial_x u + \partial_y v + \partial_z w =& 0, 
\end{align}
\end{subequations}
where ${\bf u} = (u,v,w)$ is the fluid velocity in Cartesian coordinates with position vector $\vect{x} = (x,y,z)$, $p_\textrm{tot}$ is the total pressure, $\rho_\textrm{tot}$ is the total density, and $f$ is the constant Coriolis parameter.  The boundaries are assumed to be periodic in $(x,y)$; and rigid, free-slip ($w=0$), and flat at $z=0$ and $z=-D$. Buoyancy anomalies are not permitted at the vertical boundaries in the work considered here. Note that there is no mixing allowed since equations \eqref{eqn:boussinesq} do not include viscosity or diffusivity.

\subsection{The no-motion and flattened isopycnal states}
\label{subsec:nomotionsoln}

\begin{figure}
\begin{center}
{\includegraphics[width=0.65
\textwidth]{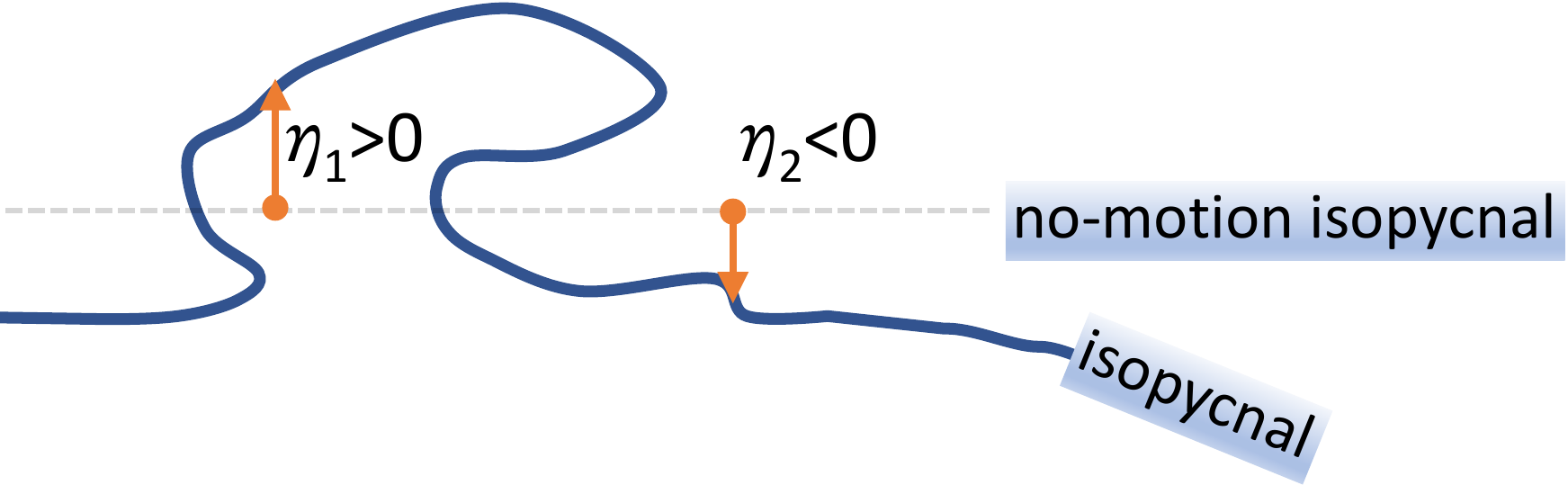}}
\caption{Schematic of adiabatic re-arrangement and isopycnal displacements. $\eta(\mathbf{x},t)$ is positive if it is above its no-motion height.}
\label{fig:Schematic}
\end{center}
\end{figure}

The most trivial, but important, solution to the equations of motion \eqref{eqn:boussinesq} is the \textit{no-motion} state which occurs when we let $\vect{u}(\vect{x},t) = (u,v,w)=(0,0,0)$. The density of the fluid $\rho_\textrm{tot}(\vect{x},t)$ must be adiabatically re-arranged to eliminate horizontal pressure and density gradients by moving parcels to their no-motions heights as sketched in Figure \ref{fig:Schematic}. The no-motion solution is denoted as $(u,v,w,p,\rho)=\left(0,0,0,-\rho_0 g z + \tilde{p}_\textrm{nm}(z),\rho_0 + \tilde{\rho}_\textrm{nm}(z) \right)$ where $\tilde{p}_\textrm{nm}$ and $\tilde{\rho}_\textrm{nm}$ are the variable part of the no-motion solution defined as
\begin{equation}
\label{nm-sol}
     \tilde{p}_\textrm{nm}(z) 
     = -g \int_0^z  \tilde{\rho}_\textrm{nm}(\xi) d\xi,
\end{equation}
so that $\partial_z \tilde{p}_\textrm{nm}(z) \equiv -g \tilde{\rho}_\textrm{nm}(z)$ and the equations of motion are satisfied.
We note that the Boussinesq approximation assumes that 
the positive constant 
$\rho_0 \gg |\tilde{\rho}_\textrm{nm}(z)|$ and $|-\rho_0gz| \gg |\tilde{p}_\textrm{nm}(z)|$.

Similar to the no-motion state, the \textit{flattened-isopycnal} state also has $w=0$ and $\rho=\rho_0 + \tilde{\rho}_\textrm{nm}(z)$. However, this state of the fluid allows for $u\neq 0$, $v\neq 0$ and $p\neq 0$, provided that $\partial_z p = 0$. As a consequence of flattening isopycnals, these states of motion have no available potential energy or vortex stretching, but do have kinetic energy and vorticity. Thus, unlike the no-motion state, a flattened isopycnal state is an achievable state-of-motion for any initial condition because energy and enstrophy can be conserved.

Importantly, the no-motion density is not the same as the mean density. These two quantities differ by a `mean density anomaly' (mda), a function of $z$ which may be quite small, or quite large (imagine an inverted fluid). This will be further discussed in section \ref{sec:mda-solution}.

\subsection{Isopycnal deviation vs excess density}
\label{subsec:iso-vs-excess}

\begin{figure}
\begin{center}
{\includegraphics[width=0.9
\textwidth]{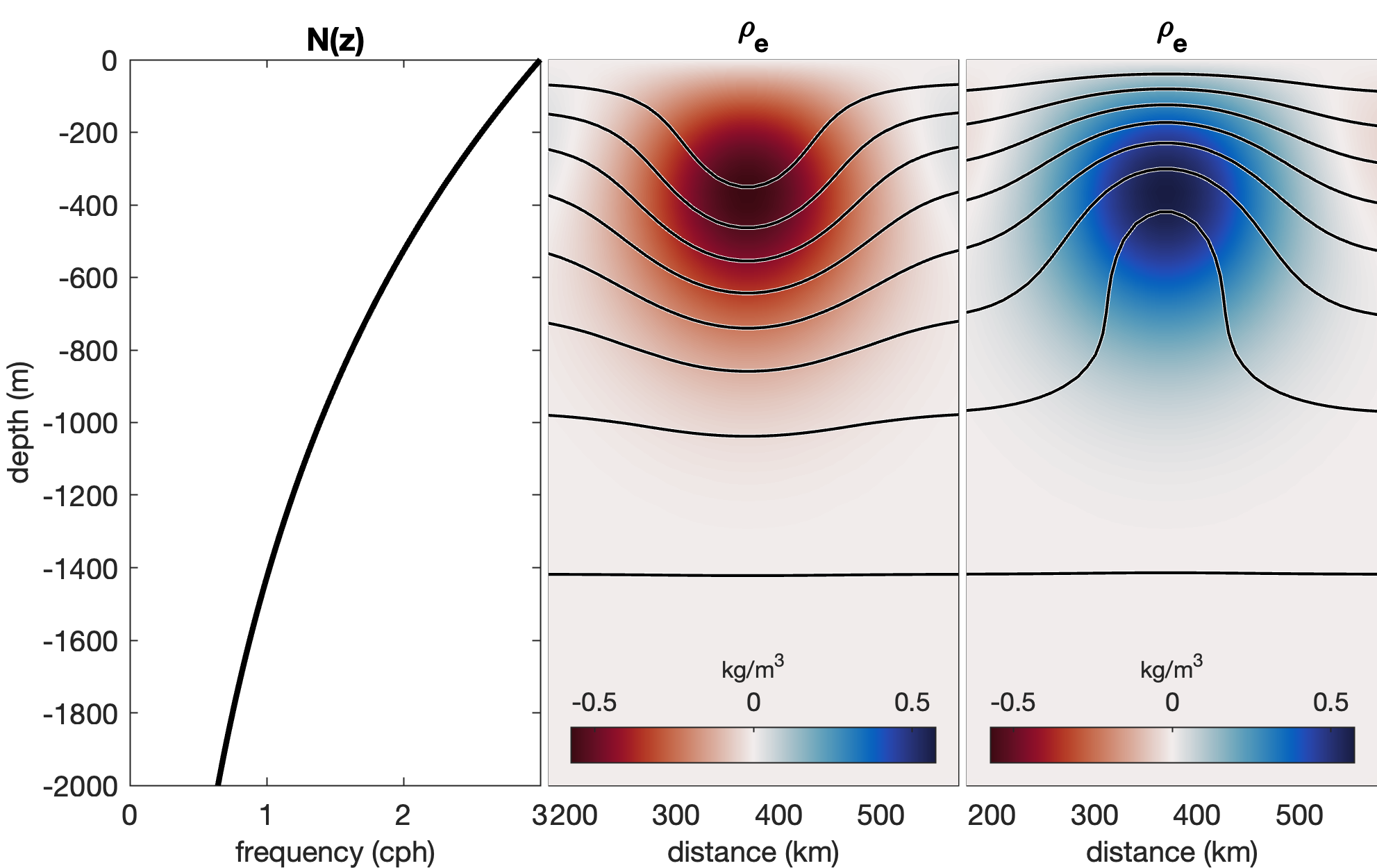}}
\caption{The buoyancy frequency of an exponential stratification profile (left) and the excess density $\rho_e$ of an anticyclonic (middle) and cyclonic eddy (right). The eddies have opposite but equal amounts of excess density (shown in color). Isopycnal lines are contoured on top.}
\label{fig:cyclonic-anticyclonic}
\end{center}
\end{figure}

There are two common approaches to expressing re-arrangements of density relative to the no-motion solution: isopycnal deviation ($\etat$) or excess density ($\rho_\textrm{e}$) defined by
\begin{subequations}
\label{iso-and-excess-def}
\begin{align}
\label{excess-def}
    \tilde{\rho}(\vect{x},t) \equiv&   \tilde{\rho}_\textrm{nm}(z) + \rho_\textrm{e}(\vect{x},t) \textrm{ or} \\
\label{iso-def}
    \tilde{\rho}(\vect{x},t) \equiv&  \tilde{\rho}_\textrm{nm}(z - {\etat}(\vect{x},t)).
\end{align}
\end{subequations}
The excess density may also be referred to as density anomaly or buoyancy anomaly where buoyancy is defined as $b\equiv -g \rho_\textrm{e}/\rho_0$. Both formulations in \eqref{iso-and-excess-def} disallow mixing in order to maintain material conservation, but do allow isopycnal overturning. Note that the argument of $\tilde{\rho}_\textrm{nm}$ in \eqref{iso-def} could be used to define an isopycnal coordinate (lines of constant $z-\etat$), but this would disallow isopycnal overturns. In practice, computing $\etat$ from $\rho_\textrm{e}$ requires inverting $\tilde{\rho}_\textrm{nm}$ in equation \eqref{iso-def} with either an analytically specified no-motion density profile, or performing an adiabatic re-arrangement of an existing density field using, e.g., equation (2.3) in \citep{winter2013-jfm}. For analytical work, it is useful to approximate the relationship between $\rho_\textrm{e}$ and $\etat$. 
Using $\tilde{\rho}_\textrm{nm}(z)$ and its inverse from the definitions in \eqref{iso-and-excess-def} results in series expansions
\begin{equation}
\begin{split}
 \etat &= - \sum_{n=1}^\infty \frac{1}{n!} \left( \frac{1}{ \partial_z \tilde{\rho}_\textrm{nm}(z)} \frac{d}{dz} \right)^{n-1} \left( \frac{1}{\partial_z \tilde{\rho}_\textrm{nm}(z)} \right)  \rho_\textrm{e}^n  \\ \rho_\textrm{e} &= 
    \sum_{n=1}^\infty \frac{(-1)^n}{n!} \tilde{\rho}^{(n)}_\textrm{nm}(z) \etat^n
\end{split}
\end{equation}
where $\partial_z \tilde{\rho}_\textrm{nm}(z) \neq 0$ and
the differential operator in the $\etat$ expansion follows from implicit differentiation of $\frac{d}{d \tilde{\rho}_\textrm{nm}}$. 
The first-order approximation to isopycnal deviation written in terms of excess density is thus given by $\etae \equiv - \rho_\textrm{e} \left( \partial_z \tilde{\rho}_\textrm{nm} \right)^{-1}$  and the first terms in the expansion are
\begin{equation}
\label{eta-expansion-explicit}
    \etat = \etae + \frac{1}{2} \partial_z \log \left( N^2(z) \right) \etae^2 + \frac{1}{12} N^4 \partial_{zz} \left( N^{-4} \right) \etae^3  + O\left( \etae^4 \right)
\end{equation}
where the squared buoyancy frequency is defined as $N^2 = -g \partial_z \tilde{\rho}_\textrm{nm}/\rho_0$. The expansion \eqref{eta-expansion-explicit} will be useful because the conservation laws for this system will all be expressed in terms of $\eta$ in section \ref{sec:mat-cons-of-z-eta}, but will need to be linearized in terms of $\etae$.  Figure \ref{fig:cyclonic-anticyclonic} shows two eddies with equal and opposite amounts of excess density, $\rho_\textrm{e}$, but very asymmetric amounts of isopycnal deviation, essentially reflecting the higher order terms from \eqref{eta-expansion-explicit}.

A key constraint is that the total density of the fluid is bounded by $\left[ \rho_0, \rho_D \right]$, where 
$\rho_0 = \rho_0 + \tilde{\rho}_\textrm{nm}(0)$ with 
$\tilde{\rho}_\textrm{nm}(0)=0$, and 
$\rho_D = \rho_0 + \tilde{\rho}_\textrm{nm}(-D)$. No fluid parcels exist with a density less than $\rho_0$ or greater than $\rho_D$, which requires
\begin{subequations}
\label{eqn:density-bounds}
\begin{align}
  0  & \leq \;  \tilde{\rho}_\textrm{nm}(z) + \rho_\textrm{e}(\vect{x},t) \leq \rho_{D} -\rho_0 \quad \textrm{or} \\
     &z \leq \;  \etat(\vect{x},t) \leq z+D.
\end{align}
\end{subequations}
Material conservation of density further requires 
\begin{subequations}
\begin{align}
\label{excess-volume-conservation}
\int \rho_\textrm{e} dV &=0 \textrm{ or} \\
\int \etat dV =& 0,
\label{iso-volume-conservation}
\end{align}
\end{subequations}
where $\int dV$ denotes integration over the entire volume of fluid. 

These two different approaches affect only the vertical momentum equation \eqref{z-momentum} and the thermodynamic equation \eqref{thermodynamic}. Starting with the vertical momentum equation \eqref{z-momentum} we subtract the no-motion solution \eqref{nm-sol} to express two forms of the vertical momentum equation, 
\begin{subequations}
\begin{align}
\ddt w
=& - \frac{1}{\rho_0} \partial_z p_\textrm{e}- \frac{1}{\rho_0} g \rho_\textrm{e}  \textrm{ or}\\
\ddt w
=& - \frac{1}{\rho_0} \partial_z p_\textrm{e}- \frac{1}{\rho_0} g \left( \tilde{\rho}_\textrm{nm}(z - {\etat})- \tilde{\rho}_\textrm{nm}(z) \right).
\end{align}
\label{eqn:w-momentum-excess}
\end{subequations}
Then to close the equations of motion, we must also express the thermodynamic equation in terms of these two quantities, e.g., 
\begin{subequations}
\begin{align} \label{thermodynamic-excess}
\ddt  \rho_\textrm{e} + 
    w\frac{d }{dz}\tilde{\rho}_\textrm{nm}(z) =& \; 0  \textrm{ or} \\\label{thermodynamic-iso}
  \ddt  {\etat}   - w =& \; 0.
\end{align}
\end{subequations}
Note that (\ref{thermodynamic-iso}) evaluated at the surface (bottom) gives
\begin{equation}
\ddt  {\etat}\big\rvert_\textrm{z=0}  = \ddt {\etat}\big\rvert_\textrm{z=-D}=0
\label{thermodynamic-iso-boundary}
\end{equation} 
since $w=0$ at the top and bottom of the domain. Relations (\ref{thermodynamic-iso-boundary}) reflect the material conservation of buoyancy anomaly at the vertical interfaces. Some literature refer to buoyancy conservation on the boundaries as a boundary condition, but this is incorrect---surface buoyancy conservation is a \emph{consequence} of the boundary condition $w=0$.

\subsection{Equations of motion}
\label{subsec:eqns-of-motion}

Any number of variations of (\ref{thermodynamic-excess})-(\ref{thermodynamic-iso}) can be used to express the equations of motion \eqref{eqn:boussinesq} relative to the no-motion solution. As we will show, the conservation laws of energy and potential vorticity are most naturally expressed in terms of isopycnal deviation $\eta(\mathbf{x},t)$, while linear solutions to the system are most naturally expressed in terms of density anomaly $ \rho_\textrm{e}(\mathbf{x},t)$.   Looking ahead towards connecting the conservation statements to the linearized equations, it is helpful to express the buoyancy as $b= -g \rho_\textrm{e}/\rho_0=-N^2 \etae $,
with total density and pressure given by,
\begin{subequations}
\label{rho-p-etae}
\begin{align}
\rho_\textrm{tot}(\mathbf{x},t) =& \rho_0 + \tilde{\rho}_\textrm{nm}(z) + \frac{\rho_0}{g} N^2 \etae(\mathbf{x},t) \\
p_\textrm{tot}(\mathbf{x},t) =& -\rho_0 g z + \tilde{p}_\textrm{nm}(z) + p_\textrm{e}(\mathbf{x},t).
\end{align}
\end{subequations}
Using \eqref{rho-p-etae}, we arrive at a particularly convenient form of the nonlinear equations:
\begin{subequations}
\label{boussinesq-tilde}
\begin{align}
\label{x-momentum-eta}
 \frac{du}{dt}  - f v  =& - \frac{1}{\rho_0} \partial_x p_\textrm{e}\\ \label{y-momentum-eta}
 \frac{dv}{dt}  + f u  =& -\frac{1}{\rho_0} \partial_y p_\textrm{e}  \\ \label{z-momentum-eta}
\frac{dw}{dt}   =& - \frac{1}{\rho_0}  \partial_z p_\textrm{e} - N^2 \etae \\ \label{thermodynamic-eta}
\frac{d}{dt} N^2 \etae - N^2 w=& 0 \\ \label{continuity-eta}
\partial_x u + \partial_y v + \partial_z w =&  0
\end{align}
\end{subequations}
with material conservation of density \eqref{excess-volume-conservation} expressed as $\int N^2 \etae \, dV = 0$.

\section{Conservation statements in terms of isopycnal displacements}
\label{sec:mat-cons-of-z-eta}

For expressing conservation statements, the primary advantage to using isopycnal deviation $z-\etat(\vect{x},t)$ instead of excess density $\rho_\textrm{e}(\vect{x},t)$ is that $z-\etat(\vect{x},t)$ is a materially conserved quantity that can be treated like a coordinate. Material conservation of $z-\etat(\vect{x},t)$ follows directly from material conservation of 
$\rho_\textrm{tot}$
\eqref{thermodynamic} and the definition of $\etat$ \eqref{iso-def}. Going further, \emph{any} quantity $f$ expressed in terms of $z-\etat(\vect{x},t)$ is materially conserved according to the equation
\begin{equation}
\label{eqn:fpe}
\ddt f(z-\etat) = 0.
\end{equation}

Following \eqref{eqn:fpe}, the volume integral of a materially conserved quantity $f(z-\etat)$ remains constant and thus must match the volume integral of any other configuration of the fluid, including the configuration where $\etat=0$, such that
\begin{equation}
\label{eqn:fpe-volume-int}
    \int f(z-\etat) dV = \int f(z) dV.
\end{equation}
This last expression says nothing about the dynamics of how the fluid moved between the different states of motions, just that fluid labels have to be conserved during such a process.

Taking advantage of \eqref{eqn:fpe} and \eqref{eqn:fpe-volume-int},
the following sections show how to arrive at expressions for available potential vorticity APV and available potential energy APE in terms of isopycnal deviation $\eta(\mathbf{x},t)$. 
Our contention is that these definitions in terms of $\eta$ are the best suited for the problem at hand, and so it is worth discussing what makes them `good' definitions. 
Any definitions of APV and APE should both appeal to our intuition, and meet the necessary dynamical requirements such as energy exchange and material conservation.  

We demand four requirements, starting with the intuitive description of APE written in Vallis \citep{vallis2006-book}: `The difference between the total potential energy of the fluid and the total potential energy after an adiabatic rearrangement to a state in which the isentropic surfaces are flat is called the available potential energy, or APE.' 
Thus, the first requirement is the {\it global requirement} that the volume integral of APE (APV) must be the difference between the volume integrals of the PE (PV) in the current and no-motion states. 
An even stronger condition is to require that the APE (APV) must vanish point-wise in the limit of no motion, which we refer to as  the {\it no motion requirement.} Third, the 
energy exchange terms in the equations for APE and KE must exactly cancel, while the material derivative of the APV must vanish, or in other words there is {\it a material requirement.} The fourth constraint demands that the linearized versions of APE (APV) must coincide with the expressions derived from the linearized equations of motion, 
{\it i.e., a linearization requirement}.

\subsection{Potential vorticity}
\label{subsec:PViso}

An important quantity characterizing (\ref{eqn:boussinesq}) is the potential vorticity 
$\Pi(\psi)$ given by 
\begin{equation}
\label{eqn:pv-background-2}
    \Pi(\psi) \equiv 
    \vect{\omega_a}\cdot \nabla \psi,
\end{equation}
where $\vect{\omega_a}=(\nabla \times \vect{u} + f\hat{z})$ is the total vorticity and  $\psi(\vect{x},t)$ is any (scalar) material invariant.  Under the Boussinesq approximation \eqref{eqn:boussinesq}, and if $\psi$ is a function of total density $\psi = \psi(\rho_\textrm{tot})$, then the potential vorticity $\Pi$ itself is also a material invariant such that
\begin{equation}
\label{eqn:Piconservatioin}
    \ddt \Pi = 0.
\end{equation}
The traditional choice $\psi$ equal to the density itself is the Ertel potential vorticity:
\begin{equation}
\textrm{Ertel PV}=  
\vect{\omega_a}\cdot \nabla \rho_\textrm{tot} = (\nabla \times \vect{u} + f\hat{z}) \cdot \nabla \rho_\textrm{tot}.
\label{ertelpv}
\end{equation}
However, Ertel PV \eqref{ertelpv} has a non-vanishing contribution 
$f \partial_z \tilde{\rho}_\textrm{nm}(z)$ for the no-motion solution, has a large Eulerian signature for internal waves (see equation 3.14 in \citep{early2021-jfm}), and linearizes to a definition inconsistent with the linearized equations of motion as discussed in Wagner \& Young \citep{wagner2015-jfm}. Thus, the traditional definition of Ertel PV potential vorticity arising from the choice $\psi= \rho_\textrm{tot}$ lacks a number of useful qualities.

On the other hand, the choice $\psi(\vect{x},t) = z-\etat(\vect{x},t) = \rho_\textrm{nm}^{-1}(\rho_\textrm{tot}(\vect{x},t))$ results in a potential vorticity 
\begin{equation}
\begin{split}
\label{eqn:pv-eta}
\textrm{PV} &\equiv 
\vect{\omega_a} \cdot \nabla (z-\etat) \\
&= 
\partial_x v - \partial_y u + f - f \partial_z \etat - (\nabla \times \vect{u}) \cdot \nabla \etat,
\end{split}
\end{equation}
with desirable features for our purposes.  
The definition \eqref{eqn:pv-eta} has
a closed-form expression that vanishes in the state of no-motion, and it linearizes to the quasigeostrophic potential vorticity which has no signature for linear internal waves.
The potential vorticity \eqref{eqn:pv-eta} is equation (107) in \citep{muller1995-rg}, although there it is  erroneously stated that overturns are not allowed. 
Notice that \eqref{eqn:pv-eta} reduces to the constant $f$ in the state of no-motion, still violating the {\it no motion requirement} for a good definition of available potential vorticity APV.  However, we next show that a good definition of APV follows from \eqref{eqn:pv-eta} and the {\it global requirement}. 

A remarkable feature of \eqref{eqn:pv-background-2} is that its volume integral over the entire fluid can be deduced entirely by surface integrals over the boundaries because of the divergence theorem and the fact that $\nabla \cdot \vect{\omega_a}=0$. Using the approach in \citep{schneider2003-jas}, we compute
\begin{subequations}
    \begin{align}
    \int \textrm{PV} \; dV &= \int \nabla \cdot \left( (z-\etat)\; \vect{\omega_a}  \right) dV \\
        &= \int (z-\etat)\; \vect{\omega_a}\cdot \hat{\vect{n}} \; dS \\
        \label{pv-surface-integral}
        &= \int f\, dV
        - \int_{z=0} \etat \left( \vect{\omega_a} \cdot \hat{\vect{z}} \right) dS + \int_{z=-D} \etat \left( \vect{\omega_a} \cdot \hat{\vect{z}} \right) dS.
    \end{align}
    \end{subequations}
To arrive at \eqref{pv-surface-integral}, we have used 
\begin{equation*}
   -D\int_{z=-D} \omega_a \cdot (-\hat {\vect{z}}) \, d S = D \int_{z=-D} (v_x -u_y+f) \, dS 
= D \int_{z=-D} f \, dS = \int f \, dV,  
\label{intermediate}
\end{equation*}
where the third equality follows from periodicity of
$(v_x-u_y)$ in the horizontal directions, such that the surface integral of $(v_x-u_y)$ is zero.  In this paper, we focus on flow in the absence of boundary buoyancy anomalies.  However, it is instructive to show how these are naturally included in the  formulation of PV and APV starting from $(z-\eta).$  Therefore we retain the terms arising from boundary buoyancy anomalies for the rest of Section \ref{subsec:PViso}.

The expression \eqref{pv-surface-integral} for PV conservation contains a constant term, $\int f \, dV$, as well as surface boundary terms $-f\int_{z=0} \eta \, dS$ and 
$f\int_{z=-D} \eta \, dS$.  Since total buoyancy at each boundary is independently conserved, the conservation of volume integrated PV may be restated as
\begin{subequations}
\begin{align}
\partial_t \int  \etat (v_x - u_y)\; dS&= 0\\
\partial_t \int  \etat \; dS&= 0,
\end{align}
\end{subequations}
with surface contributions that arise only from the top $z=0$ and bottom $z=-D.$

It is helpful to identify the range of possible values for the volume-averaged potential vorticity.  In the motionless state with no relative vorticity, the total PV from \eqref{pv-surface-integral} becomes
\begin{equation}
\label{pv-surface-integral-no-motion}
    \frac{1}{V} \int_V \Pi\, dV = f - \frac{1}{V} \int_{z=0} f \etat \, dS + \frac{1}{V} \int_{z=-D} f \etat \, dS,
\end{equation}
and thus the volume-averaged PV of the fluid can be entirely diagnosed by the buoyancy anomalies on the two boundaries. The two most extreme motionless initial conditions are the no-motion rest state (rs) and inverted, or upside-down (ud) fluid,
\begin{subequations}
\label{eqn:apv-motionless-states}
\begin{align}
\eta_\textrm{rs} &= 0 \\
\eta_\textrm{ud} &= D + 2z
\end{align}
\end{subequations}
The total PV of these two cases are $\frac{1}{V}\int_V \Pi\left[ \eta_\textrm{rs} \right]=f$ and $\frac{1}{V}\int_V \Pi\left[ \eta_\textrm{ud} \right]=-f$, values which must remain constant for all time. If the fluid is not motionless and contains relative vorticity at the boundary, the term proportional to $\int (v_x - u_y) \etat \, dS$ will not vanish and has no obvious bounds, although it too will remain constant in an adiabatic flow.

\subsection{Available potential vorticity and available potential enstrophy}
\label{subsec:APV}

To construct APV, we rewrite \eqref{pv-surface-integral} by shifting all the terms to the left-hand-side so that,
\begin{equation}
\label{apv-integral-definition}
\int \left[ \textrm{PV} - f
+ \etat \left( \vect{\omega_a} \cdot \hat{\vect{z}} \right) \delta(z) - \etat \left( \vect{\omega_a} \cdot \hat{\vect{z}} \right) \delta(z+D) \right] dV = 0
\end{equation}
and use the integrand to define available potential vorticity as $\textrm{APV} \equiv \textrm{PV}  - f + \overline{\textrm{APV}}$ where
\begin{equation}
\label{eqn:apv-eta}
\textrm{APV} \equiv \partial_x v - \partial_y u - f \partial_z \etat - (\nabla \times \vect{u}) \cdot \nabla \etat + \overline{\textrm{APV}},
\end{equation}
and
\begin{equation}
\label{eqn:mean-apv}
\overline{\textrm{APV}} \equiv  \frac{1}{A}\int_{z=0} \etat \left( \vect{\omega_a} \cdot \hat{\vect{z}} \right) dS - \frac{1}{A}\int_{z=-D} \etat \left( \vect{\omega_a} \cdot \hat{\vect{z}} \right) dS.
\end{equation}
The addition of $\overline{\textrm{APV}}$ shifts the zero point of PV such that $ \frac{1}{V}\int \textrm{APV} \, dV = 0$ for all states of motion, e.g., the rest-state and upside-down state in \eqref{eqn:apv-motionless-states}.
Material conservation of available potential vorticity APV and available potential enstrophy $\textrm{APV}^2/2$ follow directly from \eqref{eqn:Piconservatioin}, \eqref{eqn:pv-eta}, \eqref{eqn:apv-eta}-\eqref{eqn:mean-apv}, such that
\begin{equation}
    \ddt \textrm{APV} = 0,
\end{equation}
\begin{equation}
\label{eqn:enstrophy}
    \ddt \biggl (\frac{1}{2} \textrm{APV}^2
    \biggr)= 0.
\end{equation}

The idea underlying \eqref{eqn:apv-eta} was proposed by Wagner and Young \citep{wagner2015-jfm}, who describe APV as `the difference between the total PV and the PV arising by advection of the background buoyancy field.' The formulation of PV in terms of $(z-\eta)$ in \eqref{eqn:apv-eta} avoids reference to the background buoyancy, and thus leads to the closed-form expression \eqref{eqn:apv-eta} for APV, instead of a series expansion in powers of $\rho_e$ as found in \citep{wagner2015-jfm}.

The definition \eqref{eqn:apv-eta} satisfies all the requirements for a `good' definition, namely the global requirement, the no-motion requirement, the material requirement, and linearization requirement.

%
\subsection{Potential energy}
\label{sec:pe}
%

Equations (\ref{eqn:boussinesq})
globally conserve the total energy E = KE + PE, where KE and PE are the kinetic and potential energies, respectively. The equation for $\textrm{KE}=\rho_0 \vect{u}^2/2$ follows from \eqref{x-momentum}-\eqref{z-momentum},
\begin{equation}
\label{ke-full}
    \ddt \textrm{KE} = - \vect{u} \cdot \nabla \tilde{p}
    - g w \tilde{\rho},
\end{equation}
\noindent
while the equation for PE requires computing the material derivative of $\textrm{PE} = \tilde{\rho} g z,$ which after application of \eqref{thermodynamic} produces
\begin{equation}
\label{pe-full}
    \ddt  \textrm{PE} = g w \tilde{\rho}.
\end{equation}
Since the buoyancy flux term 
$w g \tilde{\rho}$ 
is opposite in sign from the 
analogous term in the kinetic energy equation \eqref{ke-full}, the sum of the two equations \eqref{ke-full} and \eqref{pe-full} results in the total energy conservation law
\begin{equation}
\label{eqn:energy-flux-total}
    \ddt  E= -\vect{u} \cdot \nabla \tilde{p}.
\end{equation}
It is noteworthy that the total energy is not materially conserved---while its depth integrated total value is constant in time, the energy of each fluid parcel includes pressure work that depends on the fluid parcel's location within the fluid.

\subsection{Available potential energy}
\label{subsec:APE}

In the original derivation of available potential energy, \citep{holliday1981-jfm} start with a kinetic energy equation using excess pressure $p_\textrm{e}$ instead of total pressure 
$p_\textrm{tot}$, that is, using (\ref{x-momentum})-(\ref{y-momentum}) with (\ref{eqn:w-momentum-excess}) instead of (\ref{x-momentum})-(\ref{z-momentum}).  Following from (\ref{eqn:w-momentum-excess}), this kinetic energy equation can be written in 2 different ways:
\begin{subequations}
\begin{align}
\ddt \textrm{KE} =& - \mathbf{u} \cdot \nabla p_\textrm{e}
    - g w \rho_\textrm{e},\; \textrm{or}\\ \label{eqn:KE-excess-eta}
    \ddt \textrm{KE} =& - \mathbf{u} \cdot \nabla p_\textrm{e}
    - g w \left( \rho_\textrm{nm}(z - {\eta}) -\rho_\textrm{nm}(z) \right).
\end{align}
\label{eqn:KE-excess}
\end{subequations}
Then to arrive at their definition of excess APE, \citep{holliday1981-jfm} perform a perturbation expansion of the term $\rho_\textrm{nm}(z - {\eta}) -\rho_\textrm{nm}(z)$. 

To arrive at the same answer, we define the materially conserved potential energy (MCPE) which is the sum of a potential term and hydrostatic pressure from each particle's no-motion position $z-\eta$.  That is, we define
\begin{equation}
\label{eqn:mcpe}
\textrm{MCPE} \equiv g (z-\eta)\tilde{\rho}_\textrm{nm} (z-\eta) + \tilde{p}_\textrm{nm}(z-\eta),
\end{equation}
where 
both terms 
are materially conserved and thus MCPE is also materially conserved. Using (\ref{eqn:fpe-volume-int}), the 
volume integral of \eqref{eqn:mcpe} leads to the equality
\begin{equation}
\label{eqn:mcpe-volume}
    \int g z \biggl[ \tilde{\rho}_\textrm{nm}(z - \eta) - \tilde{\rho}_\textrm{nm}(z) \biggr ]\, dV  = \int \biggl [ g \eta \tilde{\rho}_\textrm{nm}(z - \eta) +\tilde{p}_\textrm{nm}(z) - \tilde{p}_\textrm{nm}(z-\eta)\biggr ]\, dV,
\end{equation}
where the integrands on both sides of the equation are different forms of APE as described by the global requirement, that is, `the difference between the total potential energy of the fluid and the total potential energy after an adiabatic rearrangement to a state in which the isentropic surfaces are flat.'

Using the integrand on the right-hand-side of \eqref{eqn:mcpe-volume}, one may write APE as 
\begin{subequations}
\label{eqn:pe-hm}
\begin{align}
    \textrm{APE} &= g \eta \tilde{\rho}_\textrm{nm}(z - \eta) + \tilde{p}_\textrm{nm}(z) - \tilde{p}_\textrm{nm}(z-\eta)\\
    &= g \eta \tilde{\rho}_\textrm{nm}(z - \eta) - \int_{z-\eta}^{z} F_p\, ds\\
    &=  - \int_0^\etat g \xi \partial \tilde{\rho}_\textrm{nm}(z- \xi) d \xi,
    \label{eqn:pe-hm-c}
\end{align}
\end{subequations}
where $\mathbf{F}_p$ is the hydrostatic pressure force given by $\mathbf{F}_p = - \partial_z \tilde{p}_\textrm{nm}(z) \hat{z}$, and where the last equality follows from integration by parts and change of variables.
The relation \eqref{eqn:pe-hm} is equation (3.1) in \citep{holliday1981-jfm}, equation (2.1) in \citep{roullet2008-jfm}, and equation (3.3) in \citep{winter2013-jfm}.
As illustrated in \citep{winter2013-jfm}, the excess APE \eqref{eqn:pe-hm} is the best definition to assess the potential energy available for exchange with kinetic energy,\footnote{The definition \eqref{eqn:pe-hm} subtracts the work done by the hydrostatic pressure force $\mathbf{F}_p$ to move the particle from its no-motion position at $z-\eta$ to its current position at $z$. The additional work done by $\mathbf{F}_p$ could be included in APE to diagnose Lagrangian time series and for assessment of power.} and it
meets all four requirements for a good definition of excess available potential energy.

\subsection{Summary of Conservation Statements}
\label{sec:summary-nonlinear-conservation-laws}
%

A key feature of our methodology is that the quadratically conserved quantities in the linear system follow directly from their nonlinear counterparts. One way to achieve this goal is to 
describe the nonlinear conservation laws in terms of $z-\eta(\textbf{x},t)$, as described in Sections \ref{sec:background}-\ref{sec:pe}.

The nonlinear dynamical equations \eqref{boussinesq-tilde} have linear and quadratic conservation statements given below.  On the left are the pointwise versions of the volume-integrated statements on the right.
\begin{align}
    \nabla \cdot \mathbf{u} = 0 && \int \nabla \cdot \mathbf{u} \, dV = 0 \tag{mass} \\ 
    \frac{d}{dt} N^2 \etae = N^2 w && \int N^2 \etae \,dV = 0 \tag{potential density} \\
    \frac{d}{dt} \textrm{APV} = 0  && \int \textrm{APV} dV = 0 \tag{potential vorticity} \\
    \frac{d}{dt}   \left( \textrm{KE} + \textrm{APE} \right) = -\mathbf{u} \cdot \nabla p_\textrm{e} && \partial_t \int \left( \textrm{KE} +  \textrm{APE} \right) dV = 0 \tag{energy}
\end{align}
where
\begin{align*}
    &\textrm{KE} \equiv \rho_0 \frac{1}{2} \mathbf{u}^2 \\
    &\textrm{APE} \equiv - \int_0^\etat g \xi \partial \tilde{\rho}_\textrm{nm}(z- \xi) d \xi \\
    &\textrm{APV} \equiv \partial_x v - \partial_y u - f \partial_z \etat - (\nabla \times \vect{u}) \cdot \nabla \etat + \overline{\textrm{APV}}
\end{align*}
Notice that the conservation laws for APE and (KE + APE) are expressed in terms $\etat$.  In the next section, we will show how to linearize them together with \eqref{boussinesq-tilde}.

%
\section{Linearization}
\label{sec:linearization}
%

A key feature of our methodology is that the quadratically conserved quantities for the linear dynamical equations follow directly from their nonlinear counterparts. This requires a careful choice of variables as described in Sections \ref{sec:background}-\ref{sec:mat-cons-of-z-eta}, as well as a corresponding linear base state (here the no-motion state).

%
\subsection{Linear equations of motion 
}
\label{sec:eom-eta}
%

Linearizing \eqref{boussinesq-tilde} requires discarding any quadratic combinations of $(u,v,w,\etae,p_\textrm{e})$
\begin{subequations}
\begin{align}
\label{x-momentum-eta-lin}
\partial_t u - f v =& - \frac{1}{\rho_0} \partial_x p_\textrm{e}\\ \label{y-momentum-eta-lin}
\partial_t v + f u =& - \frac{1}{\rho_0}  \partial_y p_\textrm{e}  \\ \label{z-momentum-eta-lin}
\partial_t w  =& - \frac{1}{\rho_0}  \partial_z p_\textrm{e}
- N^2
\etae \\
\label{thermodynamic-eta-lin}
\partial_t \etae  =& w  \\ \label{continuity-eta-lin}
\partial_x u + \partial_y v + \partial_z w =& 0,
\end{align}
\label{eqn:boussinesq-eta-lin}
\end{subequations}
where \emph{all} variables are now lower order approximations of the nonlinear versions.

\subsubsection{Linearized APE}

Starting from \eqref{eqn:pe-hm-c}, one may use integration by parts to rewrite APE as
\begin{equation}
\textrm{APE}
= - \frac{g}{2} \etat^2 \partial \tilde{\rho}_\textrm{nm}(z- \etat) + \frac{g}{2}\int_0^\etat \xi^2 \partial^{(2)} \tilde{\rho}_\textrm{nm}(z- \xi) d \xi.
\end{equation}
Next, by using the relations $\tilde{\rho}_\textrm{nm}(z-\eta) = \tilde{\rho}_\textrm{nm}(z) + \rho_\textrm{e}$, $\rho_\textrm{e} = \rho_0 g^{-1} N^2 \eta_\textrm{e}$ together with the expansion
$\eta = \eta_\textrm{e} + O(\eta_\textrm{e}^2)$
(see \eqref{eta-expansion-explicit}),
one arrives at 
\begin{equation}
\textrm{APE} \sim \frac{1}{2} N^2 \etae^2 + O\left(\etae^3\right).
\end{equation}
Thus the linearized version of APE is given by $\textrm{APE} = N^2 \etae^2/2.$

\subsubsection{Linearized APV}

For the case of no boundary buoyancy anomalies considered herein, $\overline{\textrm{APV}}$ given by
\eqref{eqn:mean-apv} is zero, and hence APV reduces to
\begin{equation}
\textrm{APV} \equiv \partial_x v - \partial_y u - f \partial_z \etat - (\nabla \times \vect{u}) \cdot \nabla \etat.
\label{eqn:apv-minus-apvbar}
\end{equation}
Linearizing \eqref{eqn:apv-minus-apvbar} is achieved by discarding quadratic combinations of $(u,v,w,\eta)$ and expanding $\etat$ in terms of $\etae=-\rho_\textrm{e} \left( \partial_z \tilde{\rho}_\textrm{nm}\right)^{-1}$ using \eqref{eta-expansion-explicit}. This consistency results in
\begin{equation}
\label{eqn:qgpv-eta}
\textrm{QGPV} \equiv \partial_x v - \partial_y u - f \partial_z \etae
\end{equation}
where
\begin{equation}
    \ddt \textrm{QGPV} = 0.
\end{equation}
Then it follows that the enstrophy for the linearized dynamical equations \eqref{eqn:boussinesq-eta-lin} is defined by
\begin{equation}
\label{enstrophy-qgpv}
\mathcal{Z} \equiv \frac{1}{2} \int \textrm{QGPV}^2 dV \quad \textrm{where} \quad \partial_t \mathcal{Z} = 0.
\end{equation}

\subsubsection{Summary of linearized conservations statements}
In total, the linearized conserved quantities are
\begin{align}
    \nabla \cdot \mathbf{u} = 0 && \int \nabla \cdot \mathbf{u} \, dV = 0 \tag{mass} \\ 
    \frac{d}{dt} \etae = w && \int N^2 \etae \,dV = 0 \tag{potential density} \\
    \frac{d}{dt} \textrm{QGPV} = 0  && \int \textrm{QGPV} \,dV = 0 \tag{potential vorticity} \\
    \frac{\partial}{\partial t} \left( \textrm{KE} + \textrm{APE} \right)=  - \mathbf{u} \cdot \nabla p && \frac{\partial}{\partial t} \mathcal{E} = 0 \tag{energy} \\
    \frac{d}{dt} \textrm{QGPV}^2 = 0  &&  \frac{\partial}{\partial t} \mathcal{Z} = 0 \tag{enstrophy}
\end{align}
where
\begin{align*}
    &\textrm{KE} \equiv  \frac{1}{2} \rho_0 \vect{u}^2 \\
    &\textrm{APE} \equiv \frac{1}{2} \rho_0 N^2 \etae^2 \\
    &\textrm{QGPV} \equiv \partial_x v - \partial_y u - f \partial_z \etae \\
        & \mathcal{E} \equiv \frac{1}{2 L_x L_y}   \int \left( u^2 + v^2 + w^2 + N^2 \etae^2 \right) dV \\
    &\mathcal{Z} \equiv \frac{1}{2 L_x L_y}   \int \textrm{QGPV}^2 dV. 
\end{align*}
Each solution to the linear equations \eqref{eqn:boussinesq-eta-lin} will satisfy all of the above conservation statements. Importantly, volume-integrated conservation of mass and potential density are satisfied exactly, a requirement for physically realizable states,
which must not create mass or internal energy.

%
\subsection{Inner product space}
\label{sec:inner-product}
%

The volume integrals of 
energy $\mathcal{E}$ and enstrophy $\mathcal{Z}$ are given by 
\begin{equation}
    \mathcal{E} = \frac{1}{2 L_x L_y}  \int_{-D}^0 \int_0^{L_y} \int_0^{L_x} (u^2 +v^2 + w^2 + N^2 \etae^2) \; dV
\end{equation}
and
\begin{equation}
    \mathcal{Z} = \frac{1}{2 L_x L_y}  \int_{-D}^0 \int_0^{L_y} \int_0^{L_x} \left( \partial_x v - \partial_y u - f \partial_z \etae \right)^2\; dV.
\end{equation}
The energy takes the form of an inner-product on the sum of the squares of the 4 variables $(u,v,w,\etae)$.
The enstrophy is an inner-product on the square of one variable,  the vertical component of vorticity. Conceptually then, these two inner-products have the same general form if we imagine a two step process: first we produce an $m$-dimensional vector from the dynamical variables (e.g. velocity or vorticity), and second we dot the vector with some inner-product operator. A good notion for this idea is the bra-ket notation.


%
\subsection{Bra-ket notation}
\label{sec:bra-ket}
%


The linear equations of motion \eqref{eqn:boussinesq-eta-lin} and conservation laws can be written in bra-ket notation. Solutions of the system are represented as kets, e.g. $\ket{\psi}$, and matrix and differential operators are expressed as, e.g., $\op{H}$. We start by defining time-dependent state-vector $\ket{\psi(t)}$, such that
\begin{equation}
\label{eqn:uvw-state-vector}
\ket{\psi(t)}_{\mathcal{U}} = 
\begin{bmatrix}
    u(\vect{x},t)\\ v(\vect{x},t) \\ w(\vect{x},t)\\ \etae(\vect{x},t) \\ p_\textrm{e}(\vect{x},t)
\end{bmatrix}
\end{equation}
where $\mathcal{U} = \left\{ \ket{u(\vect{x})}, \ket{v(\vect{x})}, \ket{w(\vect{x})}, \ket{\etae(\vect{x})}, \ket{p_\textrm{e}(\vect{x})} \right\}$ is an ordered orthonormal basis for these observable flow features, such that, e.g., $u(\vect{x},t) \equiv \qprod{u(\vect{x})}{\psi(t)}$. In general we will drop the explicit dependence on $\vect{x}$ and $t$, as well as the subscript $\mathcal{U}$ until we consider a basis transformation in section \ref{sec:nonlinear-wave-vortex}. With this notation, the linear system of equations can be written as,
\begin{equation}
\label{eqn:linear-momentum-braket}
    \op{T} \ket{\psi} + \op{\Lambda} \ket{\psi} = 0
\end{equation}
where
\begin{equation}
\op{T} = 
    \begin{bmatrix}
    \partial_t & 0 & 0 & 0 & 0 \\
    0 & \partial_t & 0 & 0 & 0 \\
    0 & 0 & \partial_t & 0 & 0 \\
    0 & 0 & 0 & \partial_t & 0 \\
    0 & 0 & 0 & 0 & 0 
    \end{bmatrix}
    \quad \textrm{and} \quad 
\op{\Lambda} = 
    \begin{bmatrix}
    0 & - f & 0 & 0 & \frac{1}{\rho_0} \partial_x \\
    f & 0 & 0 & 0 & \frac{1}{\rho_0} \partial_y \\
    0 & 0 & 0 & N^2 & \frac{1}{\rho_0} \partial_z \\
    0 & 0 & -1 & 0 & 0\\
    \partial_x & \partial_y & \partial_z & 0 & 0
    \end{bmatrix}.
\end{equation}
Energy operator $\op{H}$ and enstrophy operator $\op{Z}$ are expressed as
\begin{equation}
\op{H} \equiv \frac{1}{2}
    \begin{bmatrix}
        1 & 0 & 0 & 0   & 0\\
        0 & 1 & 0 & 0   & 0\\
        0 & 0 & 1 & 0   & 0\\
        0 & 0 & 0 & N^2 & 0 \\
        0 & 0 & 0 & 0   & 0
    \end{bmatrix},
    \; \op{Z} \equiv
    \frac{1}{2}, \;
    \textrm{and} \;
\op{Q} \equiv 
    \begin{bmatrix}
        -\partial_y & \partial_x & 0 & - f \partial_z & 0
    \end{bmatrix}
\end{equation}
where $\op{Q}$ produces the vertical component of potential vorticity. With these definitions we obtain volume 
integrated energy
\begin{equation}
\label{eqn:energy-braket}
   \mathcal{E} = \braopket{\psi}{\op{H}}{\psi} = \frac{1}{2 L_x L_y}  \int_{-D}^0 \int_0^{L_y} \int_0^{L_x} (u^2 +v^2 + w^2 + N^2 \etae^2) dV.
\end{equation}
and enstrophy
\begin{equation}
\label{eqn:enstrophy-braket}
\mathcal{Z} = \braopket{\op{Q} \psi}{\op{Z}}{\op{Q}\psi} = \frac{1}{2 L_x L_y}  \int_{-D}^0 \int_0^{L_y} \int_0^{L_x} \left( \partial_x v - \partial_y u - f \partial_z \etae \right)^2 dV.
\end{equation}
It is the inner-products of energy \eqref{eqn:energy-braket} and enstrophy \eqref{eqn:enstrophy-braket} that motivate the use of bra-ket notation for this problem. When considering purely geostrophic flows, e.g. \citep{smith2013-jpo}, both inner-products can be expressed in terms of a single scalar field (e.g., a streamfunction). In the full Boussinesq flow considered here, the energy inner-product involves a 4-component vector and the enstrophy inner-product involves the scalar potential vorticity.

Linear momentum evolution in \eqref{eqn:linear-momentum-braket} can be equivalently expressed as,
\begin{subequations}
\label{eqn:braket-linear-eom}
    \begin{align}
        \op{H} \op{T} \ket{\psi} + \op{H} \op{\Lambda} \ket{\psi} =& 0 \\
        \partial_t \op{H} \ket{\psi} + \op{H} \op{\Lambda} \ket{\psi} =& 0
    \end{align}
\end{subequations}
while the linear quasigeostrophic potential vorticity evolution equation reduces to,
\begin{equation}
    \partial_t \ket{\op{Q} \psi} = 0
\end{equation}
because $\op{Q} \op{\Lambda} \ket{\psi} = 0$.

%
\section{Orthogonal solutions}
\label{sec:orthogonal-solutions}
%
\subsection{Philosophy}

Any complete basis capable of satisfying the boundary conditions of the nonlinear equations of motion can be used to represent a complete set of solutions, and is capable of representing all physically realizeable states of motion in the fluid. For example, Fourier modes in $x$, $y$ and Chebyshev polynomials in $z$ combine to form a complete basis set for each of the dynamical variables in this problem. In this scenario, the individual basis components (a Chebyshev polynomial) have no physical meaning individually, but only as a series sum representing the full solution. This is a very `numerical' solution.

At the other extreme are analytical solutions to the nonlinear equations of motion. Such solutions are individually meaningful, satisfying all boundary conditions and all conservation properties for all time (e.g., a plane-wave solution in the non-rotating version of these equations). As such, these are physically meaningful solutions from which enormous physical insight can be gained. In contrast to the more numerical approach, such solutions are often so sparse, that they are certainly not complete and thus we cannot represent all physically realizeable states of motion.

The philosophy here is to strike a balance between these two extremes---we require a solution set that is complete, capable of representing all physically realizeable states of motion, but where each basis vector also has a physically meaningful interpretation. Our approach is to find solutions to the equations of motions linearized about the no-motion state, and then require that each solution satisfy as many conservation properties as possible. We \emph{mostly} succeed at this task, falling short only with the mean-density anomaly modes, which individually do not satisfy the both constraints arising from conservation of volume-integrated potential density and conservation of volume-integrated potential vorticity (see the discussion in section \ref{sec:mda-solution}).

%
\subsection{The solutions}
%

The central idea is that any observable state-vector, $\ket{\psi}$, is composed energetically and enstrophically orthogonal observable solutions. The four major solution types, summarized in table \ref{tab:solutions}, are internal gravity waves $\ket{\wmode}$, inertial oscillations $\ket{\iomode}$, geostrophic motions $\ket{\gmode}$, and the mean-density anomaly $\ket{\mdamode}$. Thus, any observable state-vector can be expressed as,
\begin{equation}
\label{eqn:psi-expanded}
    \ket{\psi} = \sum_{k\ell j} \left( A_0^{k\ell j} \ket{\gmode} + A_\pm^{k\ell j} \ket{\wmode}  + \textrm{c.c.} \right) + \sum_{j} \left( A_-^{00j} \ket{\iomode} + \textrm{c.c.} \right) + \sum_{j} A_0^{00j} \ket{\mdamode}
\end{equation}
where $A_0^{k\ell j}$, $A_+^{k\ell j}$, and $A_-^{k\ell j}$ are time-dependent coefficient matrices. The notation here is such that the $A_0^{k\ell j}$ coefficients multiply modes with a potential vorticity signature, while the $A_\pm^{k\ell j}$ coefficients multiply wave modes with frequencies $\pm \omega_\kappa^j$. To recover a coefficient, e.g., $A_0^{k\ell j}$, we dot the state-vector $\ket{\psi}$ against an orthogonal solution and normalize, e.g.,
\begin{equation}
\label{eqn:wmode-coefficient}
    A_\pm^{k\ell j}(t) = \frac{\braopket{\wmode}{\op{H}}{\psi}}{\braopket{\wmode}{\op{H}}{\wmode}}
\end{equation}
recovers the coefficient from \eqref{eqn:psi-expanded} for the geostrophic solution $\ket{\gmode}$ by exploiting energy orthogonality. The projection operators, energy and enstrophy are summarized in table \ref{tab:solution-projection}, and detailed in the sections that follow.

That this basis is complete means that we can express the fluid state in terms of the wave-vortex eigenmodes, i.e.,
\begin{equation}
\label{eqn:wave-vortex-vector}
\ket{\psi(t) }_\mathcal{A} \equiv
    \begin{bmatrix}
        A_0^{00j}(t) \\
        A_0^{k\ell j}(t) \\
        A_-^{00j}(t) \\
        A_\pm^{k\ell j} (t)
    \end{bmatrix}
\end{equation}
where $\mathcal{A} = \left\{  \ket{\mdamode}, \ket{\gmode}, \ket{\iomode}, \ket{\wmode} \right\}$ is an ordered basis, exactly as was done in  $\eqref{eqn:uvw-state-vector}$. An additional consequence is that the combination of the forward and inverse projection form the identity matrix, analogous to writing $S S^{-1}$ for some matrix operator. In the notation here, this means that the wave-vortex projection operator $\mathcal{S}$ can be defined as
\begin{equation}
\label{eqn:wave-vortex-projection}
    \mathcal{S} \equiv \sum_{k\ell j}  \frac{ \ket{\gmode} \bra{\gmode}\op{H}}{\braopket{\gmode}{\op{H}}{\gmode}}  + \frac{ \ket{\wmode} \bra{\wmode}\op{H}}{\braopket{\wmode}{\op{H}}{\wmode}}+ \sum_{j} \frac{ \ket{\iomode} \bra{\iomode}\op{H}}{\braopket{\iomode}{\op{H}}{\iomode}} +\frac{ \ket{\mdamode} \bra{\mdamode}\op{H}}{\braopket{\mdamode}{\op{H}}{\mdamode}}
\end{equation}
such that $\mathcal{S} \ket{\psi} = \ket{\psi}_\mathcal{A}$. This will be used in section \ref{sec:nonlinear-wave-vortex} to project the equations of motion into wave-vortex space.

Once the fluid state $\ket{\psi}$ is projected onto the energetically and enstrophically orthogonal solutions, the total energy is computed with
\begin{equation}
\begin{split}
      \braopket{\psi}{\op{H}}{\psi} = &\sum_{k\ell j} 2 \left|A_0^{k\ell j} \right|^2 \braopket{\gmode}{\op{H}}{\gmode} + 2\left|A_\pm^{k\ell j} \right|^2 \braopket{\wmode}{\op{H}}{\wmode} \\ &+ \sum_{j} 2 \left|A_-^{00j}\right|^2 \braopket{\iomode}{\op{H}}{\iomode} + \left|A_0^{00j}\right|^2 \braopket{\mdamode}{\op{H}}{\mdamode}    
\end{split}
\end{equation}
and total enstrophy with,
\begin{equation}
    \braopket{\op{Q} \psi}{\op{Z}}{\op{Q}\psi} = \sum_{k\ell j} 2 \left|A_0^{k\ell j} \right|^2 \braopket{\op{Q}\gmode}{\op{Z}}{\op{Q}\gmode}  + \sum_{j}\left|A_0^{00j}\right|^2 \braopket{\op{Q}\mdamode}{\op{Z}}{\op{Q}\mdamode}
\end{equation}
where the factor of $2$ accounts for the conjugate. The solutions are eigenvectors of the operators $\op{T}$ and equivalently $-\op{\Lambda}$ such that,
\begin{equation}
    \op{T} \ket{\gmode} = 0 \ket{\gmode}, \op{T} \ket{\wmode} = \pm \omega \ket{\wmode}, \op{T} \ket{\iomode} = - f \ket{\iomode}, \op{T} \ket{\mdamode} = 0 \ket{\mdamode}
\end{equation}
where the vectors and operators are expressed relative to the basis $\left\{ \ket{u(\vect{x})}, \ket{v(\vect{x})}, \ket{w(\vect{x})}, \ket{\etae(\vect{x})} \right\}$, specifically excluding pressure from the basis $\mathcal{U}$.

\begin{table*}
\caption{Summary of energetically and enstrophically orthogonal solutions. These are the `half-complex' solutions, and must be added to their complex-conjugate for the full, physically-realizeable solution.}
\centering
\begin{minipage}{1.0\textwidth}
\begin{minipage}{1.0\textwidth}
\[
\ket{\wmode}
= \frac{1}{\omega_\kappa^j \kappa}
\begin{bmatrix}
(k\omega_\kappa^j \mp f i \ell) F^j_\kappa(z) \\
(\ell \omega_\kappa^j \pm f i k)F^j_\kappa(z) \\
- i \kappa^2 \omega_\kappa^j h_\kappa^j G^j_\kappa(z) \\
\mp \kappa^2 h_\kappa^j G^j_\kappa(z)\\
\mp \rho_0 g \kappa^2 h_\kappa^j F^j_\kappa(z)
\end{bmatrix} e^{i k x + i \ell y\pm i\omega_\kappa^j t}, \quad \ket{\iomode}
= 
\begin{bmatrix}
F^j_\textrm{io}(z) \\
i F^j_\textrm{io}(z) \\
0 \\
0\\
0
\end{bmatrix} e^{i f t}
\]
\end{minipage}
\newline
\begin{minipage}{1.0\textwidth}
\[
\ket{\gmode} =  
\begin{bmatrix}
- i \frac{g \ell}{f} F^j_\textrm{g}(z) \\
i \frac{g k}{f} F^j_\textrm{g}(z) \\
0\\
G^j_\textrm{g}(z) 
\\
\rho_0 g F^j_\textrm{g}(z)
\end{bmatrix} e^{i k x + i \ell y}, \quad
\ket{\mdamode} = 
\begin{bmatrix}
0 \\
0 \\
0\\
G^j_{\textrm{g}}(z) \\
\rho_0 g F^j_\textrm{g}(z)
\end{bmatrix}
\]
\end{minipage}
\medskip
\hrule
\end{minipage}
\label{tab:solutions}
\end{table*}


\subsection{Projection and notation}
\label{subsection-notation}

The ultimate outcome of this work are a set of orthogonal solutions and projection operators, which take the real-value observable dynamical fields $(u,v,\etae)$ and project them onto real-valued observable orthogonal solutions. The other two dynamical variables, $w$ and $p$, are diagnostically determined from the continuity equation and the vertical momentum equation. In fact, it is worth noting that the full state cannot be recovered from other combinations of the variables, only $(u,v,\etae)$ contain the complete information of the state. Thus, before even solving the equations of motion, we describe the projection operators.

The projection operators onto the eigenmode solutions are also composed of two parts: the horizontal projections, which take the form of Fourier transforms, and the vertical projections, which result from various Sturm-Liouville problems.

The solutions and projection operators require specification of $\tilde{\rho}_\textrm{nm}(z)$ and there is a strict physical requirement that $\etae$ must adhere to the condition that
\begin{equation}
     \tilde{\rho}_\textrm{nm}(z) >     \partial_z \tilde{\rho}_\textrm{nm}(z)  \eta_\textrm{e}(\mathbf{x}) > \tilde{\rho}_\textrm{nm}(z) - \tilde{\rho}_\textrm{nm}\left(-D \right),
\end{equation}
before projecting which follows from \eqref{eqn:density-bounds}. These bounds on density anomaly place strong physical limits on the vertical structure of mesoscale eddy, an idea explored in section \ref{sec:application-geostrophic-bounds}. Although not a physical requirement, for the work in this manuscript we make the additional simplifying assumption that $\etae(0)=0=\etae(-D)$. In notable contrast to $\etae$, the physical fields $(u,v)$ do not have bounds on their values.

\subsubsection{Horizontal transforms}

A simplifying assumption for this problem is that the two horizontal dimensions $(x,y)$ are periodic, which means that we can use a Fourier basis for our linear solutions. We now define the operators that move us between the spatial domain $(x,y)$ and wavenumber domain $(k,\ell)$.

The horizontal structure of all real-valued functions in the domain have the form,
\begin{equation}
    f(x,y) = \mathcal{D}^{-1}_{xy}\left[ \hat{f}(k_n, \ell_m)\right] = \hat{f}(0, 0) + \sum_{n=0}^\infty \sum_{\substack{m=-\infty\\m\neq 0}}^\infty \hat{f}(k_n, \ell_m) e^{i k_n x + i \ell_m y} + c.c
\end{equation}
where $k_n = 2 \pi n / L_x$ and $\ell_m = 2 \pi m / L_y$ and $\mathcal{D}^{-1}\left[ \cdot \right]$ is the inverse Fourier transform in two-dimensions. The Hermitian symmetry of real-valued functions allows us to assume that $c.c. = \conj{\hat{f}}(k_n, \ell_m) e^{-i k_n x - i \ell_m y}$. To recover a coefficient we define the forwards Fourier transform projection operator $\mathcal{D}\left[ \cdot \right]$,
\begin{equation}
     \hat{f}(k_n, \ell_m) = \mathcal{D}_{xy}\left[ f(x,y) \right] = \frac{1}{L_x L_y} \int_0^{L_x} \int_0^{L_y} f(x,y) e^{-i k_n x - i \ell_m y} \, dx \, dy
\end{equation}
which follows from the orthogonality, i.e., given two solution with wavenumbers $(k_a,\ell_a)$ and $(k_b,\ell_b)$
\begin{equation}
\label{eqn:fourier-orthogonality}
    \frac{1}{L_x L_y} \int_0^{L_x} \int_0^{L_y} e^{i k_a x + i \ell_a y} e^{-i k_b x - i \ell_b y} \, dx \, dy = \delta_{k_a k_b} \delta_{\ell_a \ell_b}.
\end{equation}
With these definitions, variance is preserved in both domains,
\begin{equation}
     \frac{1}{L_x L_y} \int_0^{L_x} \int_0^{L_y} f^2(x,y) \, dx \, dy = \hat{f}(0, 0) + \sum_{n=0}^\infty \sum_{\substack{m=-\infty\\m\neq 0}}^\infty 2 \left|\hat{f}(k_n,\ell_m) \right|^2.
\end{equation}
To simplify notation, anytime we write $\hat{f}$, that should be assumed to be shorthand for $\hat{f}(k_n,\ell_m) = \mathcal{D}_{xy}\left[ f(x,y) \right]$ and $\bar{f}$ is shorthand for the $k=\ell=0$ component that is simply a horizontal average.

\subsubsection{Vertical transforms}

\begin{table}
    \centering
    \begin{tabular}{l|l|l} 
         EVP & Forward transform & Inverse transform \\ \hline
        $\partial_z \left( \frac{\partial_z F_\textrm{g}}{N^2} \right) = - \frac{1}{g h_\textrm{g}} F_\textrm{g}$ & $\mathcal{F}^j_\textrm{g}[u] \equiv  \frac{1}{h_\textrm{g}^j} \int_{-D}^0 u F^j_\textrm{g} \, dz = u_\textrm{g}^j$ & $\mathcal{F}_\textrm{g}^{-1} [ u_\textrm{g}^j ]  = \sum u_\textrm{g}^j F_\textrm{g}^j= u$ \\
        $\partial_{zz} G_\textrm{g} = - \frac{N^2}{g h_\textrm{g}} G_\textrm{g}$ & $\mathcal{G}^j_\textrm{g}[\eta] \equiv \frac{1}{g} \int N^2 \eta G^j_\textrm{g} \, dz = \eta_g^j$ & $\mathcal{G}_\textrm{g}^{-1} [ \eta_g^j ] = \sum \eta_g^j G_\textrm{g}^j= \eta$ \\
        $\partial_{zz} G_\textrm{mda} = - \frac{N^2}{g h_\textrm{mda}} G_\textrm{mda}$ & $\mathcal{G}^j_\textrm{mda}[\eta] \equiv \frac{1}{g} \int N^2 \eta G^j_\textrm{mda} \, dz = \eta^j$ & $\mathcal{G}_\textrm{mda}^{-1} [ \eta^j ] = \sum \eta^j G_\textrm{mda}^j= \eta$ \\
        $\partial_{zz} G_\kappa -\kappa^2 G_\kappa = - \frac{N^2-f^2}{g h_\kappa} G_\kappa$ & $\mathcal{G}^j_\kappa[\eta] \equiv \frac{1}{g} \int \left( N^2 - f^2 \right) \eta G^j_\kappa \, dz = \eta_\kappa^j$ & $\mathcal{G}_\kappa^{-1} [ \eta_\kappa^j ] = \sum \eta_\kappa^j G_\kappa^j= \eta$ \\
        $\partial_z \left( \frac{\partial_z F_\textrm{io}}{N^2 - f^2} \right) = - \frac{1}{g h_\textrm{io}} F_\textrm{io}$ & $\mathcal{F}^j_\textrm{io} \left[ u \right] \equiv \frac{1}{h_\textrm{io}^j} \int u F^j_\textrm{io} \, dz = u_\textrm{io}^j$ & $\mathcal{F}_\textrm{io}^{-1} [ u_\textrm{io}^j ]  = \sum u_\textrm{io}^j F_\textrm{io}^j= u$ 
    \end{tabular}
    \caption{The first and second column show the vertical mode projection operators and their inverses. In this notation, the function $u$, $\eta$, $F$, $G$ and $N$ are all functions of $z$. All eigenmodes have the relationship $F = h \partial_z G$, but $N^2 G_\textrm{g} = -g \partial_z F_\textrm{g}$ for the geostrophic solution and $(N^2 - \omega_\kappa^2) G_\kappa = -g \partial_z F_\kappa$ for the internal gravity wave solution.}
    \label{tab:vertical-mode-projection}
\end{table}

To solve the linear system (\ref{eqn:boussinesq-eta-lin}), it is helpful to notice that (\ref{x-momentum-eta-lin})-(\ref{y-momentum-eta-lin}) imply the same vertical structure for $(u,v,p)$, while (\ref{thermodynamic-eta-lin}) requires the same vertical structure for $(w,\etae)$. In general, the vertical structure will be wavenumber dependent, and we will use $F^{k\ell j}(z)$ for the vertical structure of $\hat{u}^{k\ell}(z),\hat{v}^{k\ell}(z),\hat{p}^{k\ell}(z)$ and $G^{k\ell j}(z)$ for the vertical structure of $\hat{w}^{k\ell}(z),\hat{\etae}^{k\ell}(z)$. For compactness of notation throughout the rest of section \ref{sec:orthogonal-solutions}, we will use $\partial_z$ to denote ordinary derivatives with respect to $z$.  We will also abbreviate $F^{k\ell j}(z), G^{k\ell j}(z)$ as 
$F^{j}, G^{j}$ except when we need to specify special horizontal wavenumbers.

 All solution types
 will satisfy a Sturm-Liouville ordinary differential equation of the form
\begin{equation}
\partial_z (p(z) \partial_z\phi(z)) + q(z) \phi(z) = - \frac{\sigma(z)}{g h^j} \phi(z),\quad -D < z < 0,
\label{eqn:sturm-liouville}
\end{equation}
together with separated boundary conditions, where $\phi(z)$ is either $G^j(z)$ or $F^j(z)$, and the Sturm-Liouville eigenvalue is $(gh^j)^{-1}.$ The specific boundary conditions and functions $p(z) > 0$ and $\sigma(z) > 0$ depend on the solution type (see table \ref{tab:vertical-mode-projection}).  

There are other noteworthy features of the vertical structure functions $F^j(z)$ and $G^j(z)$.  In particular, all solutions satisfy $F^j(z) = h^j \partial_z G^j(z)$ (although the relation between $G^j(z)$ and $\partial_z F^j(z)$ is different for different solution types, given in 
table \ref{tab:vertical-mode-projection}). Furthermore, for all solution types, orthogonality may be expressed in terms of $G^j(z)$, given by 
\begin{equation}
    \int_{-D}^0 \sigma(z) G^i(z) G^j(z) \, dz = g\delta_{ij}
\end{equation}
with either $\sigma(z)=N^2(z)$ or $\sigma(z)=N^2(z) - f^2$, depending on the particular problem.

Just as with the Fourier transform, we will \emph{always} be able to define forward and inverse transforms $\mathcal{G}$ and $\mathcal{G}^{-1}$ for the $G^j(z)$ modes,
\begin{subequations}
    \begin{align}
    \eta(z) =& \mathcal{G}^{-1} [ \eta^j ] =  \sum_{n=0} \eta^j G^j(z) \\
    \eta^j =& \mathcal{G}^j[\eta(z)] = \int_{-D}^0 \sigma(z) \eta(z) G^j(z) \, dz
    \end{align}
\end{subequations}
and \emph{sometimes} be able to define similar projection operators $\mathcal{F}$ and $\mathcal{F}^{-1}$ for the $F^j(z)$ modes. For geostrophic solutions, there is an eigenvalue problem for $F^j(z)$, indicating that we can partition the horizontal kinetic energy independent of the total energy.  However, in the case of wave solutions, there is no Sturm-Liouville problem for 
$F^j(z)$,
indicating that the horizontal kinetic energy cannot be separately partitioned.

\subsection{Geostrophic projection}

\begin{table}
    \centering
    \begin{tabular}{l|l|l|l} 
         Projection & Energy $\braopket{\vect{\Psi}}{\op{H}}{\vect{\Psi}}$& Enstrophy $\qprod{\op{Q}\vect{\Psi}}{\op{Q}\vect{\Psi}}$ & Range \\ \hline
         $A_0^{k\ell j} = - \frac{f}{g} \frac{1}{\kappa^2 + \lambda_j^{-2}} \left( \mathcal{F}^j_g \left[ \hat{\zeta} \right] - \frac{f}{h_g^j} \mathcal{G}^j_g [ \hat{\eta} ]  \right)$ & $\frac{g}{2}   \left(\kappa^2 + \lambda_j^{-2} \right)\lambda_j^2$ & $\frac{g}{2} \left( \kappa^2  + \lambda_j^{-2} \right)^2 \lambda_j^2$ & $k > 0, j \geq 1$ \\
         $A^{kl0}_0 = - \frac{f}{g \kappa^2} \mathcal{F}^j_g \left[ \hat{\zeta} \right]$ & $\frac{g}{2} \kappa^2 \lambda_0^2$ & $\frac{g}{2}  \kappa^4 \lambda_0^2 $ & $k > 0, j = 0$ \\
         $A_0^{00j} = \mathcal{G}^j_\textrm{mda} [ \bar{\eta}_\textrm{e} ] = \frac{1}{\rho_0 g}\mathcal{F}^j_\textrm{mda} [ \bar{p}_\textrm{e} ]$ & $\frac{g}{2}$ & $\frac{g}{2} \lambda_j^{-2}$ & $k = 0, j \geq 1$ \\
         $A_\pm^{k\ell j} =  \frac{e^{\mp i \omega_\kappa^j t}}{2 \kappa h_\kappa^j}  \left( i \mathcal{G}^j_\kappa \left[\hat{w}(z) \right] \mp \omega_\kappa^j \mathcal{G}^j_\kappa \left[\hat{\eta} - \hat{\eta}_g \right] \right)$ & $h_\kappa^j$ & $0$ & $k > 0, j \geq 1$ \\
         $A_-^{00j} = \frac{e^{-i f t}}{2} \mathcal{F}^j_\textrm{io} \left[ \bar{u} - i  \bar{v}\right]$ & $h_\textrm{io}^j$ & $0$ & $k = 0, j \geq 0$
    \end{tabular}
    \caption{Projection operators, energy, enstrophy, and valid range for the five primary solution types. These projection operators take physical variables $(u,v,\etae)$ and transform them into wave-vortex space.}
    \label{tab:solution-projection}
\end{table}

The geostrophic solutions have vertical modes $F_\textrm{g}^j(z)$ for the vertical structure of $u$ and $v$, which follows from the eigenvalue problem
\begin{equation}
\label{eqn:f-evp-geostrophic}
    \partial_z\left( \frac{f^2}{N^2} \partial_zF^j_\textrm{g}\right) = - \frac{f^2}{g h^j_\textrm{g}} F^j_\textrm{g}
\end{equation}
with boundary conditions $\partial_z F^j_\textrm{g}(0) = 0 = \partial_z F^j_\textrm{g}(-D)$. As a regular Sturm-Liouville problem the eigenmodes satisfy the orthogonality condition,
\begin{equation}
\label{eqn:f-norm-geostrophic}
    \int_{-D}^0 F_\textrm{g}^i(z) F_\textrm{g}^j(z) \, dz = \gamma^j \delta_{ij} \quad \textrm{where} \quad \gamma^j = \begin{cases}
        D & j=0 \\
        h^j_\textrm{g} & \textrm{otherwise}
    \end{cases}
\end{equation}
with chosen normalization $\gamma^j$, and thus can be used to define projection operator $\mathcal{F}_\textrm{g}[u]$ and its inverse $\mathcal{F}^{-1}_\textrm{g}[u_g^j]$ as defined in the first row of table \ref{tab:vertical-mode-projection}. The vertical structure of $\etae$ is described by $G^j_\textrm{g} = - \frac{g}{N^2} \partial_z F^j_\textrm{g}$ which also forms an eigenvalue problem
\begin{equation}
\label{eqn:g-evp-geostrophic}
 \partial_{zz}G^j_\textrm{g} = - \frac{N^2}{g h^j_\textrm{g}} G^j_\textrm{g}
\end{equation}
with boundary conditions $G^j_\textrm{g}(0)=0=G^j_\textrm{g}(-D).$ The normalization of the orthogonality condition for the $G_\textrm{g}$ modes
\begin{equation}
\label{eqn:g-norm-geostrophic}
    \int_{-D}^0 N^2(z) G_\textrm{g}^i(z) G_\textrm{g}^j(z) \, dz = g \delta_{ij}
\end{equation}
must be determined by from \eqref{eqn:f-evp-geostrophic} using integration-by-parts. The projection operator $\mathcal{G}_\textrm{g}[\eta]$ and its inverse $\mathcal{G}^{-1}_\textrm{g}[\eta_g^j]$ follow from \eqref{eqn:g-evp-geostrophic} and are defined in the second row of table \ref{tab:vertical-mode-projection}. 

It will prove useful to note that the relationships
\begin{subequations}
\label{eqn:gmodes-diff-int}
    \begin{align}
        \mathcal{F}^j_g \left[ \frac{\partial \eta}{\partial z} \right] =&  \frac{1}{h_g^j} \mathcal{G}^j_g \left[ \eta \right] \\
        \mathcal{G}^j_g \left[ \frac{1}{N^2}\frac{\partial u}{\partial z} \right] =& - \frac{1}{g} \mathcal{F}^j_g [ u ]
    \end{align}
\end{subequations}
can be used to integrate and differentiate variables.

To determine the coefficients $A_0^{k\ell j}$ of each geostrophic solution from observed values $(u,v,w,\eta)$, we project onto the geostrophic solution. In practice, this looks like
\begin{subequations}
\label{a0-projection}
    \begin{align}
       \braopket{\gmode}{\op{H}}{\vect{\psi}}  =&  \frac{1}{2} \int_
        {-D}^0\left( i \frac{g}{f}l F_g^j(z) \hat{u} - i \frac{g}{f}k F_g^j(z) \hat{v}  + N^2 G_g^j(z) \hat{\eta} \right) dz  \\
        =&   \frac{1}{2} \left( - i \frac{g h_g^j}{f} \mathcal{F}^j_g \left[ \hat{\zeta} \right] + g \mathcal{G}^j_g [ \hat{\eta} ]  \right)
    \end{align}
\end{subequations}
where we have defined the vertical component of vorticity with
\begin{equation}
    \hat{\zeta} \equiv i k \hat{v} - i l \hat{u}.
\end{equation}
The total energy of this solution follows,
\begin{subequations}
\label{a0-energy}
    \begin{align}
       \braopket{\gmode}{\op{H}}{\gmode}  =& \frac{1}{2} \int_
        {-D}^0\left( \frac{g^2}{f^2} (l^2 + k^2) F_g^j(z) F_g^i(z)  + N^2 G_g^j(z) G_g^i(z) \right) dz \\
        =& \frac{1}{2}  g \left(\kappa^2 \lambda_j^2  + 1 \right) \delta^{ij}
    \end{align}
\end{subequations}
where we have expressed the eigenvalue $h^j_\textrm{g}$ as the squared deformation radius, $\lambda_j^2 \equiv \frac{g h^j_\textrm{g}}{f^2}$. Combining \eqref{a0-projection} with \eqref{a0-energy} means that the coefficient $A_0^{k\ell j}$ is recovered with,
\begin{equation}
A_0^{k\ell j} \ket{\gmode}= \ket{\gmode} \frac{\braopket{\gmode}{\op{H}}{\vect{\psi}}}{\braopket{\gmode}{\op{H}}{\gmode}}= - \frac{f}{g} \frac{1}{\kappa^2 + \lambda_j^{-2}} \left( \mathcal{F}_g \left[ \hat{\zeta} \right] - \frac{f}{h_g^j} \mathcal{G}_g [ \hat{\eta} ]  \right) \ket{\gmode}.
\end{equation}
This projection operator, as well as energy and enstrophy for these modes, are shown in the first row of table \ref{tab:solution-projection}.

The EVP \eqref{eqn:f-evp-geostrophic} admits a $j=0$ mode, $F_\textrm{g}^0(z)=1$, with eigenvalue $\frac{f^2}{g h^0_g} = 0$, known as the barotropic mode. This mode has no density anomaly, $G_\textrm{g}^0(z)=0$, and thus has zero APE and no vortex stretching. As a mode with only vorticity, it is rightfully called a `vortical' mode. Its projection follows exactly the same methodology using normalization $\lambda_0^2 = \frac{gD}{f^2}$ and is shown as the second row of table \ref{tab:solution-projection}. Although this mode is simply a geostrophic solution, because of its special characteristics it is denoted separately as $\ket{\gOmode}$.

\subsection{Mean density anomaly projection}
\label{sec:mda-solution}

The mean density anomaly (mda) solution is the horizontally averaged ($\kappa=0$) solution with zero-frequency $\omega=0$.  The mda solution has $(u, v, w) = 0$, and thus the only equation of motion that remains is the horizontally-averaged vertical momentum equation,
\begin{equation}
    N^2 \bar{\eta}_e = - \frac{1}{\rho_0} \partial_z \bar{p}_e,
\label{eqn:mda-constraint}
\end{equation}
where the bar denotes the horizontal average. The mda solutions are exactly the difference between the average density in the fluid, $\bar{\rho}(z)$, and the no-motion density, $\rho_\textrm{nm}(z),$ as noted in section \ref{subsec:nomotionsoln}. This mode was not included in \citep{early2021-jfm} because there $\bar{\rho}(z)$ was used to define the background state whereas here we use $\rho_\textrm{nm}(z)$.\footnote{An anonymous reviewer of \citep{early2021-jfm} first brought this mode to our attention.} However, the presence of the mean density anomaly is important for constructing a complete basis when linearizing about the no-motion state. Simply swapping two fluid parcels in the water column results in an mda solution and changes the potential energy and enstrophy of the system.

Depth-integrating \eqref{eqn:mda-constraint},
one finds that
the mda solutions are constrained by global conservation of potential density
\begin{equation}
\label{eqn:mda-density-conservation}
\int_{-D}^{0} N^2 \bar{\eta}_e dz = -\int_{-D}^{0}\frac{1}{\rho_0} \partial_z \bar{p}_e = \bar{p}_e(0) - \bar{p}_e(-D)=0,
\end{equation}
and they are also constrained by global conservation of QGPV,
\begin{equation}
\label{eqn:mda-qgpv-conservation}
\int_{-D}^{0} \partial_z \bar{\eta}_e dz = \bar{\eta}_e(0) - \bar{\eta}_e(-D) = 0.
\end{equation}
In the rigid-lid problem, \eqref{eqn:mda-density-conservation} tells us that the average pressure anomaly at the two boundaries must be zero, with $\bar{p}_e(-D)=\bar{p}_e(0)=0.$ In addition, \eqref{eqn:mda-qgpv-conservation} says that the average buoyancy anomaly at the boundaries must vanish, with  $\bar{\eta}_e(-D)=\bar{\eta}_e(0)=0$.

Even though the mda solutions have no fluid velocity, their energy and enstrophy are both nonzero, and the $G_\textrm{g}(z)$ eigenmodes found from \eqref{eqn:g-evp-geostrophic} diagonalize both quantities. Thus it is natural to describe the mda solutions using \eqref{eqn:g-evp-geostrophic} together with appropriate boundary conditions
that also satisfy the constraints 
 \eqref{eqn:mda-density-conservation}-\eqref{eqn:mda-qgpv-conservation}. 
 However, if we impose both $p_\textrm{e}(-D)=p_\textrm{e}(0) = 0$ \emph{and} $\eta_\textrm{e}(-D)=\eta_\textrm{e}(0)=0$, then the boundary value problem is overdetermined. 
 Therefore we must ask: what is the correct choice?

Enforcing \eqref{eqn:mda-density-conservation} with $p_\textrm{e}(-D)=p_\textrm{e}(0)=0$ on the geostrophic eigenvalue problem \eqref{eqn:g-evp-geostrophic} results in physically realizeable mda solutions, while imposing \eqref{eqn:mda-qgpv-conservation} with $\eta_\textrm{e}(-D) = \eta_\textrm{e}(0) = 0$ is more practical for idealized numerical simulations. If the zero pressure anomaly conditions are used, then the resulting solutions will have have a non-zero buoyancy anomaly at the boundary. In the most general Boussinesq model this is a physically realizeable state of the system.

Disallowing boundary buoyancy anomalies and restricting to the special case $\eta_\textrm{e}(-D) = \eta_\textrm{e}(0) = 0$ (i.e.\ $G(-D)=G(0)=0$), we must disregard the lack of physical realizeability for each individual mda mode, because we will not be imposing global conservation of potential density (global conservation of internal energy) given by \eqref{eqn:mda-density-conservation}.
With this choice, we must settle for realizability of the total mda solution, $\sum_j A_0^{00j} \ket{\mdamode}$. On the other hand, the choice $G(-D)=G(0)=0$ allows us to re-use the geostrophic modes, which has a number of obvious practical advantages.

Proceeding with the choice $G(-D)=G(0)=0$, projection of $\bar{\eta}_\textrm{e}$ onto the $G_\textrm{g}(z)$ eigenmodes follows immediately. Using integration by parts, it can be shown that
\begin{equation}
    \mathcal{G}_\textrm{g}[\bar{\eta}_\textrm{e}] =  \frac{1}{\rho_0 g}\mathcal{F}_\textrm{g}[\bar{p}_\textrm{e}],
\end{equation}
as noted in the third row of table \ref{tab:solution-projection}, where we denote the coefficients of the mda solution $\ket{\mdamode}$ as $A_0^{00j}$.

It is quite easy to unintentionally violate the pressure and buoyancy anomaly constraints \eqref{eqn:mda-density-conservation}-\eqref{eqn:mda-qgpv-conservation} with an invalid initial condition, and thus any numerical model should check these constraints upon initialization, after which time they will continue to be imposed by the dynamics. In section \ref{sec:application-geostrophic-bounds} we consider a geostrophic streamfunction designed to simultaneously satisfy both constraints. 

\subsection{IGW projection}

The required relationships between the vertical modes $F_\kappa$ (for $u$, $v$, and $p$) and $G_\kappa$ (for $w$ and $\etae$) are
\begin{equation}
        F_\kappa = h_\kappa \partial_z G_\kappa \textrm{ and } (N^2- \omega_\kappa^2) G_\kappa = -g\partial_z F_\kappa
\end{equation}
which follow from the continuity equation and the vertical momentum equation, respectively. These relationships lead to the eigenvalue problem \begin{equation}
\label{eqn:evp-g-igw}
    \partial_{zz} G_\kappa^j - \kappa^2 G_\kappa^j = - \frac{N^2-f^2}{g h_\kappa^j} G_\kappa^j
\end{equation}
with boundary conditions $G_\kappa(0)=0=G_\kappa(-D)$, where we used that
\begin{equation}
\omega_\kappa^j \equiv \sqrt{ g h^j_\kappa\kappa^2 + f^2}.
\end{equation}
The quantity $\omega_\kappa^j$ should be viewed as a shorthand for the wave frequency that is dependent on the eigenvalue $h_\kappa^j$ and the total wavenumber $\kappa = \sqrt{k^2 + \ell^2}$. 
Unlike constant stratification or hydrostatics, the eigenvalue problem for $\omega$ does not lead to a complete basis of orthogonal solutions.

Vertical modes $G_\kappa$ therefore satisfy the orthogonality condition
\begin{equation}
\label{eqn:g-norm-igw}
    \frac{1}{g} \int_{-D}^0 \left( N^2 - f^2\right) G_\kappa^i G_\kappa^j \, dz = \delta_{ij}
\end{equation}
which leads to projection operator $\mathcal{G}^j_\kappa[\eta]$ and its inverse $\mathcal{G}^{-1}_\kappa[\eta_w^j]$ as defined in the fourth row of table \ref{tab:vertical-mode-projection}. Unlike the geostrophic modes, there is no eigenvalue problem associated with the $F_\kappa$ modes as discussed in \citep{early2021-jfm}.

To determine the wave coefficients $A_\pm^{k\ell j}$ we again exploit orthogonality and project the observed state $\ket{\psi}$ onto the internal gravity wave mode. This particular calculation is more challenging and we save the details for appendix \ref{appendix:wave-mode-projection}. The net result is
\begin{equation}
\label{eqn:igw-projection}
A_\pm^{k\ell j} = \frac{\braopket{\wmode}{\op{H}}{\vect{\psi}}}{\braopket{\wmode}{\op{H}}{\wmode}}= \frac{e^{\mp i \omega_\kappa^j t}}{2 \kappa h_\kappa^j}  \left( i \mathcal{G}^j_\kappa \left[\hat{w} \right] \mp \omega_\kappa^j \mathcal{G}^j_\kappa \left[\hat{\eta} - \hat{\eta}_g \right] 
        \right)
\end{equation}
where $\hat{\eta}_g=A_0^{k\ell j}$ is the isopycnal deviation of the geostrophic mode at this same wavenumber. This solution projection operator is shown in the fourth row of table \ref{tab:solution-projection}. The vertical velocity can be replaced in favor of the horizontal divergence, such that
\begin{equation}
    A_\pm^{k\ell j} =  \frac{e^{\mp i \omega_\kappa^j t}}{2 \kappa h_\kappa^j}  \left( -i \mathcal{G}^j_\kappa \left[\mathcal{G}_g^{-1} \left[ h_g^j \mathcal{F}^j_g \left[  i k \hat{u} + i l \hat{v} \right] \right] \right] \mp \omega_\kappa^j \mathcal{G}^j_\kappa \left[\hat{\eta} - \hat{\eta}_g \right] \right),
\end{equation}
where we have used the relationships in \eqref{eqn:gmodes-diff-int} and the continuity equation, $-\partial_z \hat{w} = i k \hat{u} + i l \hat{v}$.

\subsection{IO projection}

The inertial oscillations exist at the IGW limit where $\kappa=0$ and $\omega=f$, such that $G_\textrm{io}$ satisfies
\begin{equation}
\label{eqn:io-g-evp}
    \partial_{zz} G^j_\textrm{io} = - \frac{N^2-f^2}{g h^j_\textrm{io}} G^j_\textrm{io},
\end{equation}
with $G^j_\textrm{io}(0) = 0 =G^j_\textrm{io}(-D)$. 
The corresponding boundary value problem for $F^j_\textrm{io}$ is given by
\begin{equation}
\label{eqn:io-f-evp}
    \partial_z \left( \frac{\partial_z F^j_\textrm{io}}{N^2 - f^2} \right) = - \frac{1}{g h^j_\textrm{io}} F^j_\textrm{io},
\end{equation}
with $\partial_z F^j_\textrm{io}(0)=0=\partial_z F^j_\textrm{io}(-D).$
Thus for the inertial oscillations with $\kappa=0$, orthogonality of $F^j_\textrm{io}$ may be stated as
\begin{equation}
    \int_{-D}^0 F_\textrm{io}^i F_\textrm{io}^j \, dz = h_\textrm{io}^i \delta^{ij},
\end{equation}
as indicated in the last row of table \ref{tab:vertical-mode-projection}. 

Projection onto the inertial oscillation solution follows with,
\begin{subequations}
    \begin{align} 
    \braopket{\iomode}{\op{H}}{\vect{\psi}}  =& \frac{e^{-i f t}}{2} \int_{-D}^0 \left( \bar{u} - i  \bar{v} \right) F_\textrm{io}(z) dz \\
    =& \frac{e^{-i f t}}{2} h_\textrm{io}^j \mathcal{F}^j_\textrm{io} \left[ \bar{u} - i  \bar{v}\right]
    \end{align}
\end{subequations}
which means that,
\begin{equation}
A_-^{00j} = \frac{\braopket{\iomode}{\op{H}}{\vect{\psi}}}{\braopket{\iomode}{\op{H}}{\iomode}}= \frac{e^{-i f t}}{2} \mathcal{F}^j_\textrm{io} \left[ \bar{u} - i  \bar{v}\right]
\end{equation}
as shown in the last row of table \ref{tab:solution-projection}.

\subsection{Geostrophic streamfunction projection}

If the fluid state can be described entirely in terms of a geostrophic streamfunction, the projection onto the geostrophic modes simplifies. A geostrophic streamfunction is proportional to the pressure anomaly $\psi_g \equiv \frac{1}{\rho_0 f} p_\textrm{e}$ where then $u=-\partial_y \psi_g$, $v=\partial_x \psi_g$ and $N^2 \etae = - f \partial_z \psi_g$ or, equivalently, $\rho_\textrm{e} = -  \frac{\rho_0 f}{g} \partial_z \psi_g$. This means that the projection operator can now be written as,
\begin{subequations}
\label{eqn:streamfunction}
    \begin{align}
        A_0^{k\ell j} =& - \frac{f}{g} \frac{1}{\kappa^2 + \lambda_j^{-2}} \left( \mathcal{F}^j_g \left[ \hat{\zeta} \right] - \frac{f}{h_g^j} \mathcal{G}^j_g [ \hat{\eta} ]  \right) \\
        =& - \frac{f}{g} \frac{1}{\kappa^2 + \lambda_j^{-2}} \left( - \kappa^2 \mathcal{F}^j_g \left[ \hat{\psi}_g \right] + \frac{f^2}{h_g^j} \mathcal{G}^j_g \left[ \frac{1}{N^2 }\partial_z \hat{\psi}_g \right]  \right) \\
        =& - \frac{f}{g} \frac{1}{\kappa^2 + \lambda_j^{-2}} \left( - \kappa^2 \mathcal{F}^j_g \left[ \hat{\psi}_g \right] - \frac{f^2}{g h_g^j} \mathcal{F}^j_g [ \hat{\psi}_g ]  \right) \\
        =& \frac{f}{g} \mathcal{F}^j_g [ \hat{\psi}_g ],
    \end{align}
\end{subequations}
where we used the relation from \eqref{eqn:gmodes-diff-int} in the third step.

\subsection{Summary of solutions}
\label{sec:summary-of-solutions}

These four solutions fall into two obvious categories: 1) wave solutions which include the internal gravity waves $\ket{\wmode}$ and inertial oscillations $\ket{\iomode}$ and 2) potential vorticity (or vortex) solutions which include geostrophic motions $\ket{\gmode}$ and the mean-density anomaly $\ket{\mdamode}$. The key feature separating these solution types being whether or not they have potential enstrophy.

There is 
another division of solution types: solutions with and without flattened isopycnals. In particular the inertial oscillation $\ket{\iomode}$ and barotropic geostrophic $\ket{\gOmode}$ solutions are flattened-isopycnal solutions. In section \ref{subsec:nomotionsoln} we noted that both the no-motion solution and the flattened-isopycnal solutions have no density anomaly, but that the flattened-isopycnal solutions are a physically realizeable state-of-motion from any given initial conditions. What this means in practice is that, given some initial state of motion $\ket{\psi_0}$ with energy $\mathcal{E}_0$ and enstrophy $\mathcal{Z}_0$, it is possible to construct a new state $\ket{\psi_\textrm{flat}}$ entirely from the basis $\left\{ \ket{\iomode}, \ket{\gOmode} \right\}$ with the same with energy $\mathcal{E}_0$ and enstrophy $\mathcal{Z}_0$.

%
\section{Nonlinear equations of motion and energy fluxes}
\label{sec:nonlinear-wave-vortex}
%

Inserting the wave-vortex projection operator/identity operator $\mathcal{S}$ from \eqref{eqn:wave-vortex-projection} into the linear momentum equations reduces to the almost trivial statement
\begin{equation}
    \left( \partial_t  + \op{\Lambda} \right) \mathcal{S} \ket{\psi}_\mathcal{U} = \partial_t \ket{\psi}_\mathcal{A} = 0,
\end{equation}
which tells us that the coefficients of the linear eigenmodes are constant in time. 
For example, the equation for the linear evolution of the inertia-gravity  waves is given by
\begin{equation}
    \left( \partial_t  + \op{\Lambda} \right) \frac{ \ket{\wmode} \braopket{\wmode}{\op{H}}{\psi} }{\braopket{\wmode}{\op{H}}{\wmode}} = A_\pm^{k\ell j}  \left( \partial_t  + \op{\Lambda} \right) \ket{\wmode} + \partial_t A_\pm^{k\ell j} \ket{\wmode} = \partial_t A_\pm^{k\ell j}  \ket{\wmode},
\end{equation}
where we used \eqref{eqn:wmode-coefficient} and $\left( \partial_t  + \op{\Lambda} \right) \ket{\wmode}=0$ from \eqref{eqn:braket-linear-eom}.

Modifying the linearized equations of motion \eqref{eqn:linear-momentum-braket} to include the nonlinear terms from \eqref{boussinesq-tilde} leads to
\begin{equation}
\label{eqn:nonlinear-eom-braket}
    \left( \partial_t  + \op{\Lambda} \right) \ket{\psi}_\mathcal{U} + \left[\op{\Lambda}^{NL}  \ket{\psi} \right]_\mathcal{U}  = 0,
\end{equation}
where
\begin{equation}
\label{eqn:nonlinear-operator}
    \left[\op{\Lambda}^{NL}  \ket{\psi} \right]_\mathcal{U} = 
    \begin{bmatrix}
    \textrm{uNL} \\
    \textrm{vNL} \\
    0 \\
    \textrm{nNL} \\
    0
    \end{bmatrix} \equiv 
    \begin{bmatrix}
        u \partial_x u + v \partial_y u + w \partial_z u \\
        u \partial_x v + v \partial_y v + w \partial_z v \\
        0 \\
        u \partial_x \etae + v \partial_y \etae + w \left( \partial_z \etae + \etae \partial_z \ln N^2 \right)\\
        0
    \end{bmatrix}
\end{equation}
is expressed in terms of the ordered basis $\mathcal{U}$. To project into wave-vortex space we again apply $\mathcal{S}$ from \eqref{eqn:wave-vortex-projection} so that \eqref{eqn:nonlinear-eom-braket} becomes
\begin{equation}
    \partial_t \ket{\psi}_\mathcal{A} = - \mathcal{S} \op{\Lambda}^{NL}  \ket{\psi}
\end{equation}
which is explicitly,
\begin{equation}
\label{eqn:nonlinear-wave-vortex-eqn}
\partial_t
    \begin{bmatrix}
        A_0^{00j} \\
        A_0^{k\ell j} \\
        A_-^{00j} \\
        A_\pm^{k\ell j} 
    \end{bmatrix}
    = -
    \begin{bmatrix}
         \mathcal{G}^j_\textrm{mda} [ \overline{\textrm{nNL}} ] \\
         - \frac{f}{g} \frac{1}{\kappa^2 + \lambda_j^{-2}} \left( \mathcal{F}^j_g \left[ \hat{Z} \right] - \frac{f}{h_g^j} \mathcal{G}^j_g [ \hat{N} ]  \right) \\
         \frac{e^{-i f t}}{2} \mathcal{F}^j_\textrm{io} \left[ \overline{\textrm{uNL}} - i  \overline{\textrm{vNL}}\right] \\
         \frac{e^{\mp i \omega_\kappa^j t}}{2 \kappa h_\kappa^j}  \left( i \mathcal{G}^j_\kappa \left[\hat{\Delta}(z) \right] \mp \omega_\kappa^j \mathcal{G}^j_\kappa \left[\hat{N} - \hat{N}_g \right] \right) 
    \end{bmatrix}
\end{equation}
where
\begin{subequations}
    \begin{align}
        \hat{Z}\equiv& i k \mathcal{D}_{xy}\left[ \textrm{vNL} \right] - i l \mathcal{D}_{xy}\left[ \textrm{uNL} \right] \\
        \hat{\Delta}\equiv& \mathcal{G}_g^{-1} \left[ h_n \mathcal{F}_g \left[ i k \mathcal{D}_{xy}\left[ \textrm{uNL} \right] + i l \mathcal{D}_{xy}\left[ \textrm{vNL} \right] \right] \right] \\
        \hat{N} \equiv& \mathcal{D}_{xy}\left[ \textrm{nNL} \right].
    \end{align}
\end{subequations}
This is presented in `pseudospectral' form, where the nonlinear terms are expressed in terms of $(u,v,\etae,w)$ and are then transformed using the projection operators from table \ref{tab:solution-projection}. Of course, the components $(u,v,\etae,w)$ in \eqref{eqn:nonlinear-operator} could be further expressed in terms of their constituent parts, e.g., $u=u_\textrm{g} + u_\textrm{w} + u_\textrm{io}$.

\subsection{Stratified quasigeostrophic potential vorticity equation}

Restricting the basis to only geostrophic modes, \eqref{eqn:nonlinear-wave-vortex-eqn} reduces to
\begin{multline}
\label{eqn:model-qgpv}
    - (\kappa^2 + \lambda_j^{-2}) \frac{g}{f} \partial_t A_0^{k\ell j} + \mathcal{F}^j_g \left[ \mathcal{D}_{xy} \left[ \left( u_g \partial_x + v_g \partial_y\right) \left( \partial_x v_g - \partial_y u_g \right) \right] \right] \\ - \frac{f}{h_g^j} \mathcal{G}^j_g [ \mathcal{D}_{xy} \left[ \left( u_g \partial_x + v_g \partial_y\right) \eta_g \right] ] = 0.
\end{multline}
Using \eqref{eqn:gmodes-diff-int}, \eqref{eqn:model-qgpv} further reduces to
\begin{equation}
    - (\kappa^2 + \lambda_j^{-2}) \frac{g}{f} \partial_t A_0^{k\ell j} + \mathcal{F}^j_g \left[ \mathcal{D}_{xy} \left[ \left( u_g \partial_x + v_g \partial_y\right) \left( \partial_x v_g - \partial_y u_g - f \partial_z \eta_g\right) \right] \right]  = 0,
\end{equation}
or simply
\begin{equation}
\label{eqn:qgpve}
    \left( \partial_t +  u_g \partial_x + v_g \partial_y\right) \left( \partial_x v_g - \partial_y u_g - f \partial_z \eta_g\right) = 0.
\end{equation}
This is, of course, the stratified quasigeostrophic potential vorticity equation (qgpve) which could equivalently be expressed in terms of a streamfunction using \eqref{eqn:streamfunction}. From this perspective then, the qgpve model \eqref{eqn:qgpve} is an \emph{aliased} version of the full Boussinesq equations \eqref{eqn:nonlinear-wave-vortex-eqn}. Energy that should have gone into other modes is instead added back into the geostrophic modes.

Many other reduced interaction models are possible and are useful for isolating important dynamics \citep{hernandez2014-jfm,hernandez2021-jpo}.




%
\section{Applications}
\label{sec:applications}
%

Here we show some of the consequences of the work presented here. We start by demonstrating how isopycnal deviation is computed from excess density using exponential stratification as an example. Using a geostrophic streamfunction for an eddy, we then show how strict adherence to the boundary conditions places strong limits on the vertical structure of mesoscale eddies. These bounds imply an asymmetry between cyclonic and anticyclonic eddies. Finally, we show how the APV defined here captures the height nonlinearity found in shallow-water PV. The result of this nonlinearity is another source of asymmetry between cyclonic and anticyclonic eddies.


\subsection{Isopycnal deviation in exponential stratification}

Consider the `canonical' Garrett-Munk exponential stratification profile
\begin{equation}
\label{eqn:exp-stratification}
    \rho_\textrm{exp}(z) = \rho_0 + \frac{\rho_0}{g} \frac{b}{2} N_0^2 \left(1 - \exp\left( \frac{2z}{b}\right)\right)
\end{equation}
where $N_0 = 3$ cycles per day, $b=1300$ m, in an ocean of depth $D=4000$ m \citep{munk1981-book}. An important quantity will be the total change in density in the fluid, $\Delta \rho \equiv \rho_D - \rho_0$,
\begin{equation}
    \Delta \rho_\textrm{exp} = \frac{\rho_0}{g} \frac{b}{2} N_0^2 \left(1 - \exp\left( -\frac{2 D}{b}\right)\right) \approx \rho_0 \frac{b N_0^2}{2 g}.
\end{equation}
Given some observed density state $\rho_\textrm{tot}({\bf x},t)$, excess density is simply $\rho_\textrm{e}({\bf x},t) = \rho_\textrm{tot}({\bf x},t)- \rho_\textrm{nm}(z)$, straight from \eqref{excess-def}. Computing $\eta({\bf x},t)$ requires we invert the stratification profile of the no-motion state, so let us define a functional from the definition of $\rho^{-1}_\textrm{nm}$, denoted as $\mathcal{Z}$, that maps density back to its no-motion configuration, i.e.,
\begin{equation}
\mathcal{Z}_\textrm{exp} \left[ \rho_\textrm{tot}({\bf x},t) \right] = \frac{b}{2} \ln \left(1 + \frac{2g}{bN_0^2} \left( 1 - \frac{\rho_\textrm{tot}({\bf x},t)}{\rho_0} \right) \right),
\end{equation}
where $\eta({\bf x},t) = z - \mathcal{Z} \left[ \rho_\textrm{tot}({\bf x},t) \right]$ from \eqref{iso-def}. With this, we now have an explicit relationship between excess density and isopycnal deviation,
\begin{equation}
\label{eqn:eta-exp-stratification}
    \eta({\bf x},t) = z - \mathcal{Z}_\textrm{exp} \left[ \rho_\textrm{exp} (z) + \rho_\textrm{e}({\bf x},t) \right] = - \frac{b}{2} \ln \left( 1 - \frac{2}{b} \etae \right).
\end{equation}
The isopycnal deviation is simply a re-scaling of excess density in the constant stratification case, the same reason that linearized Ertel PV and QGPV also coincide for constant stratification. However, for exponential stratification, the relationship between the two quantities is not so simple.


\subsection{Asymmetric bounds of a shallow geostrophic eddy}
\label{sec:application-geostrophic-bounds}

\begin{figure}
\begin{center}
{\includegraphics[width=0.45
\textwidth]{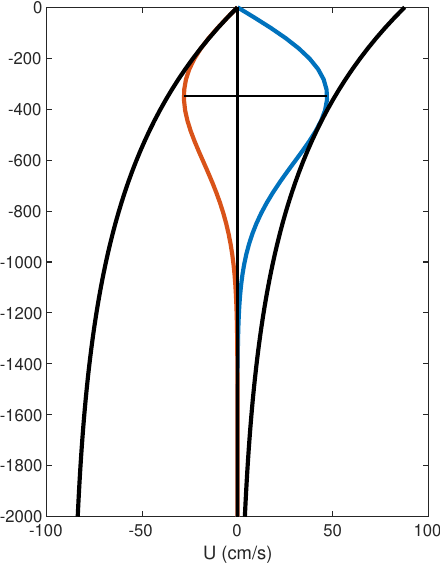}}
\caption{The black lines show the minimum and maximum allowable density anomaly amplitude, expressed in terms fluid velocity at the surface, for an eddy with $L_e=120$ km and $H_e=350$ m. The red line shows the maximum allowable anticyclonic amplitude, $U\approx 30$ cm/s, while the blue line shows the maximum allowable cyclonic amplitude $\approx 45$ cm/s.}
\label{fig:eddy-bounds}
\end{center}
\end{figure}

Consider a geostrophic streamfunction for a mesoscale eddy of the form,
\begin{equation}
    \psi(\mathbf{x}) = \frac{U L_e e^{\frac{1}{2}}}{\sqrt{2}} e^{-\frac{z^2}{2 H_e^2}} \left[  e^{-\frac{r^2}{L_e^2} } -  \frac{\pi L_e^2}{L_x L_y}  \right]
    \label{eqn:psi-eddy}
\end{equation}
where $U$ is the maximum fluid velocity, $L_e$ is the horizontal length scale, $H_e$ is the depth of the maximum buoyancy anomaly, and $r^2=x^2 + y^2$. Figure \ref{fig:cyclonic-anticyclonic} shows a vertical cross-section of the excess density of two eddies with $H_e=300$ m, $L_e=120$ km, and $U=\pm30$ cm/s in exponential stratification following \eqref{eqn:exp-stratification}. Potential density conservation, $\int N^2 \etae \, dV=0$, is equivalent to requiring that $\int_A \psi(x,y,0) \, dA = 0$ and $\int_A \psi(x,y,-D) \, dA = 0$. The second term on the right hand side of \eqref{eqn:psi-eddy} ensures that this condition is met at $z=0$. However, the scale $H_e$ controls magnitude of the pressure anomaly at the bottom boundary, and thus we require that $\psi\left(0,0,-D\right) / \psi\left(0,0,-H_e \right) < 10^{-p}$, i.e.,
\begin{equation}
    H_e < \frac{D}{\sqrt{p \ln 10 + \frac{1}{2}}}
\end{equation}
which means that $H_e \approx \frac{D}{4}$ is the approximate longest allowable scale for this case with no surface buoyancy anomalies if we take $p=7$ (the number of significant digits in single precision floating point).

A consequence of enforcing potential density conservation is that the far-field sea-surface height (away from the eddy) is not zero. The sea-surface height is given by
\begin{equation}
    \textrm{ssh} =\frac{f}{g} \psi(x,y,0)= \frac{f U L_e}{g} \frac{e^{\frac{1}{2}}}{\sqrt{2}}  \left[  e^{-\frac{r^2}{L_e^2} } -  \frac{\pi L_e^2}{L_x L_y}  \right],
\end{equation}
with maximum at $(x,y,z)=(0,0,0)$ and minimum at $(x,y,z)=(L_x,L_y,0)$, the latter of which is proportional to $\frac{L_e^2}{L_x L_y}$. A positive surface pressure anomaly at the eddy center results from a negative density beneath (warmer, lighter fluid), but because the lighter fluid had to come from somewhere and displace the heavier fluid, this results in a negative surface pressure anomaly away from the eddy.

The density anomaly for the eddy is
\begin{equation}
\rho_\textrm{e}(\mathbf{x}) =  - \frac{f \rho_0}{g} \frac{\partial \psi}{\partial z} = \frac{f \rho_0}{g} \frac{U L_e e^{\frac{1}{2}}}{\sqrt{2}} \frac{z}{H_e^2} e^{-\frac{z^2}{2 H_e^2}} \left[  e^{-\frac{r^2}{L_e^2} } -  \frac{\pi L_e^2}{L_x L_y}  \right]
\end{equation}
where the boundary conditions considered here require that $\rho_\textrm{e}(0)=0$ and $\rho_\textrm{e}(-D)=0$, conditions automatically met provided $H_e < \frac{D}{4}$. Using \eqref{eqn:eta-exp-stratification} we can explicitly compute the isopycnal deviation,
\begin{equation}
\label{eqn:eta-eddy}
    \eta({\bf x}) =  - \frac{b}{2} \ln \left( 1 - \frac{\sqrt{2} f U L_e e^{\frac{1}{2}}}{b N_0^2 H_e^2} z e^{-\frac{2z}{b}-\frac{z^2}{2 H_e^2}} \left[  e^{-\frac{r^2}{L_e^2} } -  \frac{\pi L_e^2}{L_x L_y}  \right]\right).
\end{equation}

Next we need to establish the maximum allowable amplitude such that the eddy does not exceed the density bounds. Starting from \eqref{eqn:density-bounds} evaluated at $(x,y)=(0,0)$, we require that
\begin{equation}
    -\tilde{\rho}_\textrm{nm}(z) \leq  \rho_\textrm{e}(0,0,z) \leq \Delta \rho - \tilde{\rho}_\textrm{nm}(z) 
\end{equation}
which means that
\begin{equation}
    - \gamma^{-1} \frac{ N_0^2 b }{\sqrt{2} f } \frac{H_e}{L_e}  \left(1 - e^{\frac{2z}{b}} \right) \leq U \frac{z}{H_e} e^{- \frac{z^2}{2 H_e^2} + \frac{1}{2}} \leq \gamma^{-1} \frac{ N_0^2 b }{\sqrt{2} f } \frac{H_e}{L_e}  \left( e^{\frac{2z}{b}} - e^{-\frac{2 D}{b}}\right),
\end{equation}
where
\begin{equation}
    \gamma \equiv 1 -  \frac{\pi L_e^2}{L_x L_y}.
\end{equation}
For this particular vertical structure, these bounds are exact. However, we can go a step further and approximate these bounds as,
\begin{equation}
\label{eqn:eddy-bounds}
    \frac{ N_0^2 b }{\sqrt{2} f } \frac{H_e}{L_e} \frac{H_e}{2 b} e^{-\frac{1}{2}} \gtrapprox U \gtrapprox - \frac{ N_0^2 b }{\sqrt{2} f } \frac{H_e}{L_e} \left( e^{-\frac{2 z_0}{b}} - e^{-\frac{2D}{b}} \right),
\end{equation}
which is valid for small $H_e$.

Figure \ref{fig:eddy-bounds} shows the bounds for an eddy with $L_e=120$ km and $H_e=350$ m. Notable is that the bounds are asymmetric, and dependent on the location in the water column. The anticyclonic eddy shown in figure \ref{fig:cyclonic-anticyclonic} is at its maximum allowable amplitude, while the cyclonic eddy can have its amplitude increased by another 50\% per figure \ref{fig:eddy-bounds}.

These results show that cyclonic eddies near the surface can have substantially higher amplitudes than anticyclonic eddies, while the deep eddies have more symmetric bounds and even favor anticyclones for the largest values of $H_e = \frac{D}{4}$. This asymmetry is consistent with geophysical turbulence experiments showing a strong skew towards cyclonic vortices near the upper-boundary and even a skew towards anticylones mid-water column \citep{roullet2010-prl}.

These bounds may have consequences for how we interpret the interior structure of large, mesoscale eddies \citep{chelton2011-pio}. The eddy bounds \eqref{eqn:eddy-bounds} scale with $\frac{H_e^2}{L_e}$ and $\frac{H_e}{L_e}$ and place strong restrictions on the vertical structure and amplitude of mesoscale eddies where $L_e \sim O \left(100 \textrm{km} \right)$, thus limiting the range of physically allowable vertical structures differently for cyclones and anticyclones. However, when $L_e$ is relatively small, then the range of allowable amplitudes is far beyond what is observed in the ocean.

\subsection{The height nonlinearity in APV}

APV expressed in terms of isopycnal deviation $\eta$ closely resembles potential vorticity from the shallow-water equations, thereby allowing for congruent dynamical reasoning. This is particularly useful for the height nonlinearity which arises in finite-amplitude flows, and adds a dynamical correction to quasigeostrophy important for mesoscale ocean features \citep{anderson1979-dsr, charney1981-book}. Here we first derive this correction in shallow-water using an approach related to \citep{petviashvili1980-jetp}, and then show the analogous argument with APV.

In the shallow-water equations the total depth of the layer is given by $h=D+\eta_s$, and when combined with absolute vorticity $\omega_a=\zeta^z +f,$ results in a potential vorticity conservation equation \citep{remmel2009-jfm},
\begin{equation}
\label{eqn:apv-sw-conservation}
    \frac{d}{dt} \left( \frac{\omega_a}{h} \right) = 0.
\end{equation}
Defining shallow-water available potential vorticity as $\textrm{APV}_\textrm{sw} = D \frac{\omega_a}{h} - f$, the shallow-water version of QGPV is recovered by expanding
\begin{equation}
\textrm{APV}_\textrm{sw} = \left(\zeta^z +f \right) \left( 1 + \frac{\eta_s}{D} \right)^{-1} - f
\end{equation}
and assuming $\eta_s/D \sim O(\epsilon)$ and $\zeta^z/f \sim O(\epsilon)$ so that $\textrm{QGPV}_\textrm{sw} \equiv \zeta^z - f \frac{\eta_s}{D}$ after neglecting terms of $O(\epsilon^2)$. However, it is quite often the case that height ratio $\eta_s/D$ is not nearly as small as the Rossby number, which means that
\begin{equation}
\label{eqn:apv-sw-expansion}
\textrm{APV}_\textrm{sw} \approx \zeta^z - f \frac{\eta_s}{D} + f \left( \frac{\eta_s}{D} \right)^2 + f O( \epsilon^3)
\end{equation}
if we take $\eta_s/D \sim O(\epsilon)$ and $\zeta^z/f \sim O(\epsilon^2)$.


If we express APV \eqref{eqn:apv-eta} in terms of $\etae$ and take $\partial_z \etae \sim O(\epsilon)$, $\etae \partial_z \log N^2 \sim O(\epsilon)$, and $\zeta^z/f \sim O(\epsilon^2)$, then
\begin{equation}
\label{eqn:apv-general-expansion}
\textrm{APV} \approx  \zeta^z  - f \partial_z \etae  - f \frac{1}{2} \partial_z \log N^2 \, \partial_z \eta_e^2 + f  O(\epsilon^3).
\end{equation}
The quadratic height term in \eqref{eqn:apv-general-expansion} is analogous to the quadratic height term in shallow-water \eqref{eqn:apv-sw-expansion}, but with a dependence on stratification and location in the water column. This is a generalization of the same effect noted in \citep{anderson1979-dsr} and it does not occur in constant stratification.

\begin{figure}
\begin{center}
{\includegraphics[width=0.9
\textwidth]{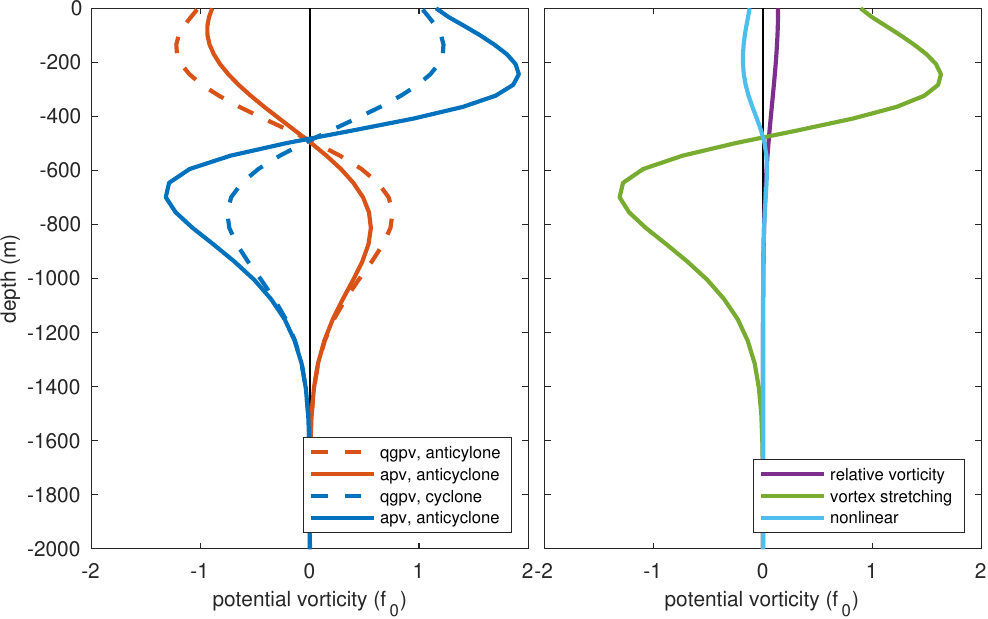}}
\caption{The left panel shows the QGPV (dashed line) and APV (solid line) through the center of the cyclonic (blue) and anticyclonic (red) eddy. The right panel shows the relative vorticity (purple), vortex stretching (green), and nonlinear terms (light blue), of the cyclonic eddy.}
\label{fig:pv-comparison}
\end{center}
\end{figure}

For the shallow geostrophic eddy example considered here, the APV in exponential stratification can be explicitly computed using \eqref{eqn:eta-exp-stratification}, resulting in
\begin{equation}
\textrm{APV}_\textrm{exp} = \partial_x v - \partial_y u - \left( f \partial_z \etae + (\nabla \times \vect{u}) \cdot \nabla \etae \right) \left(1 - 2\frac{\etae}{b} \right)^{-1},
\end{equation}
which shows the vortex stretching and nonlinear term multiplied by the same factor $(1-2 \etae/b)^{-1}$. This quantity is nearly $O(1)$, as excursions of $\etae$ are large compared to the scale depth $b$. Figure \ref{fig:pv-comparison} shows that the vortex stretching term is, by an order of magnitude, the most significant term for these mesoscale eddies. The nonlinear height correction to vortex stretching is more important than either the relative vorticity or the nonlinear term, and additionally causes an asymmetry in cyclones and anticyclones not present in QGPV.



%
\section{Discussion}
\label{sec:discussion}
%

An interpretation of the wave-vortex decomposition as fundamentally about `PV-inversion' emerges from the form of the projection operator for internal-gravity waves (fourth row in table \ref{tab:solution-projection}). To separate the internal gravity waves from the rest of the flow, one must \emph{first} compute the potential vorticity and deduce the associated geostrophic streamfunction (a PV-inversion \citep{hoskins1985-qjrms}), and only then can the internal gravity wave structures be deduced. In the case of constant stratification or hydrostatic flows, the vertical mode projection operators all commute, and this fact is quickly obscured in the derivation. At least conceptually then, it is helpful to think about the wave-vortex decomposition as originating from a definition of potential vorticity.

Perhaps one of the most significant contributions of this work is the closed-form definition of available potential vorticity (APV) in terms of isopycnal deviation. Given the interpretation of the wave-vortex decomposition as a PV-inversion, this definition of APV allows one to determine the order of approximation for each mode in the decomposition and thus assess the fidelity of the decomposition.

In summary, the work presented here continues the march towards increasing realism in wave-vortex decompositions by including the effects of arbitrary stratification, and fully articulating the relationship between the linear decomposition on the nonlinear available potential vorticity from which it is derived.

\section*{Acknowldgements} Leslie Smith gratefully acknowledges the support by the Deutsche Forschungsgemeinschaft (DFG) through the Research Unit FOR5528. J. Early's work was supported by NSF grants OCE-2123740 and OCE-2048552 as well as NASA award 80NSSC21K1823. G. H-D was supported, in part, by grants UNAM-DGAPA-PAPIIT IN112222 and Conahcyt A1-S-17634. M.P. Lelong was supported by NSF OCE-2123740. G. H-D would like to thank the hospitality of NorthWest Research Associates and the support of UNAM-PASPA-DGAPA during his sabbatical visit.

\begin{appendices}

\section{Wave mode projection}\label{appendix:wave-mode-projection}

The wave mode projection operator is derived by first noting that because $\braopket{\wmode}{\op{H}}{\gmode}=0$, we can write that $\braopket{\wmode}{\op{H}}{\vect{\psi}}=\braopket{\wmode}{\op{H}}{\vect{\psi}-\gmode}$. In the derivation that follows, we drop all superscripts and subscripts from $\omega_\kappa^j$ and $h_\kappa^j$. The Fourier transformed total field $(\hat{u},\hat{v},\hat{w},\hat{\eta})$, Fourier transformed geostrophic fields $(\hat{u}_g,\hat{v}_g,\hat{\eta}_g)$ as well as $F_\kappa$ and $G_\kappa$ are all functions of $z$. For shorthand we define $\hat{\zeta} \equiv i k \hat{v} - i \ell \hat{u}$ and $\hat{\delta} \equiv i k \hat{u} + i \ell \hat{v}$.
\begin{align*} 
    & \braopket{\wmode}{\op{H}}{\vect{\psi}} \\
     = & \braopket{\wmode}{\op{H}}{\vect{\psi}-\gmode} \\\nonumber
       =&   \frac{e^{\mp i \omega t}}{2\omega \kappa} \int_
        {-D}^0\left( \left( k \omega \pm i f \ell \right) F_\kappa \left( \hat{u} - \hat{u}_g \right) + \left( \ell \omega \mp i f k \right) F_\kappa \left( \hat{v} - \hat{v}_g \right) + i \kappa^2 \omega h G_\kappa \hat{w} \mp  N^2 \kappa^2 h G_\kappa(z) \left( \hat{\eta} - \hat{\eta}_g \right) \right) dz \\  
        =&   \frac{e^{\mp i \omega t}}{2\omega \kappa} \int_{-D}^0\left( -i\omega \hat{\delta} F_\kappa  + i \kappa^2 \omega h G_\kappa \hat{w} \mp  f \left(\hat{\zeta}-\hat{\zeta}_g \right) F_\kappa \mp  N^2 \kappa^2 h G_\kappa(z) \left( \hat{\eta} - \hat{\eta}_g \right) \right) dz \\
        =& \frac{e^{\mp i \omega t}}{2} \left[ \frac{i}{\kappa} \mathcal{G}_\kappa \left[\hat{w}(z) \right] \mp \frac{1}{\omega \kappa}  \int_{-D}^0\left(   f \left(\hat{\zeta}-\hat{\zeta}_g \right) F_\kappa +  N^2 \kappa^2 h G_\kappa(z) \left( \hat{\eta} - \hat{\eta}_g \right) \right) dz \right] \\
        =& \frac{e^{\mp i \omega t}}{2} \left[ \frac{i}{\kappa} \mathcal{G}_\kappa \left[\hat{w}(z) \right] \mp \frac{1}{\omega \kappa}  \int_{-D}^0\left(   f \left(\hat{\zeta}-\hat{\zeta}_g \right) F_\kappa +  \frac{1}{g} N^2 \left( \omega^2 - f^2 \right) G_\kappa(z) \left( \hat{\eta} - \hat{\eta}_g \right) \right) dz \right] \\ \nonumber
        =& \frac{e^{\mp i \omega t}}{2}  \left[ \frac{i}{\kappa} \mathcal{G}_\kappa \left[\hat{w}(z) \right] \mp \frac{1}{\omega \kappa}  \int_{-D}^0\left(   f \left(\hat{\zeta}-\hat{\zeta}_g \right) F_\kappa +\frac{f^2}{g}  \left( \omega^2  - N^2 \right) G_\kappa(z) \left( \hat{\eta} - \hat{\eta}_g \right)  \right) dz \right] \\
        & + \frac{e^{\mp i \omega t}}{2}  \left[  \frac{1}{\omega \kappa}  \int_{-D}^0\left(   \frac{\omega^2}{g} \left(N^2-f^2\right) G_\kappa(z) \left( \hat{\eta} - \hat{\eta}_g \right) \right) dz \right] \\
        =&  \frac{e^{\mp i \omega t}}{2}  \left[ \frac{i}{\kappa} \mathcal{G}_\kappa \left[\hat{w}(z) \right] \mp \frac{\omega}{\kappa} \mathcal{G}_\kappa \left[\hat{\eta} - \hat{\eta}_g \right] \mp \frac{1}{\omega \kappa}  \int_{-D}^0\left(   f \left(\hat{\zeta}-\hat{\zeta}_g \right) F_\kappa + f^2 \partial_z F_\kappa(z) \left( \hat{\eta} - \hat{\eta}_g \right)  \right) dz \right] \\
        =& \frac{e^{\mp i \omega t}}{2}  \left[ \frac{i}{\kappa} \mathcal{G}_\kappa \left[\hat{w}(z) \right] \mp \frac{\omega}{\kappa} \mathcal{G}_\kappa \left[\hat{\eta} - \hat{\eta}_g \right] \mp \frac{f}{\omega \kappa}  \int_{-D}^0\left(    \left(\hat{\zeta}-\hat{\zeta}_g \right) - f \partial_z  \left( \hat{\eta} - \hat{\eta}_g \right)  \right) F_\kappa(z) dz
        \right] \\
        =& \frac{e^{\mp i \omega t}}{2}  \left[ \frac{i}{\kappa} \mathcal{G}_\kappa \left[\hat{w}(z) \right] \mp \frac{\omega}{\kappa} \mathcal{G}_\kappa \left[\hat{\eta} - \hat{\eta}_g \right] 
        \right]
\end{align*}
where we used that 
\begin{equation}
\mathcal{G}_\kappa \left[\hat{w}(z) \right] =   \int_{-D}^0 \left( h \kappa^2 \hat{w}(z) G_\kappa-\hat{\delta}(z) F_\kappa \right)\, dz
\end{equation}
in the third step which follows from
\begin{subequations}
    \begin{align}
        \int_{-D}^0 \hat{\delta}(z) F_\kappa \, dz=& \int_{-D}^0 \hat{w}(z) \partial_z F_\kappa \, dz \\
        =& - \frac{1}{g} \int_{-D}^0 \left( N^2 - \omega^2 \right) \hat{w}(z) G_\kappa \, dz \\
        =& - \frac{1}{g} \int_{-D}^0 \left( N^2 - f^2 \right) \hat{w}(z) G_\kappa \, dz + h \kappa^2 \int_{-D}^0 \hat{w}(z) G_\kappa \, dz \\
        =& - \mathcal{G}_\kappa \left[\hat{w}(z) \right] + h \kappa^2 \int_{-D}^0 \hat{w}(z) G_\kappa \, dz.
    \end{align}
\end{subequations}
The total energy is,
\begin{subequations}
    \begin{align} 
    \braopket{\wmode}{\op{H}}{\wmode}  =& \frac{1}{2} \left[ \frac{i}{\kappa} \mathcal{G}^j_\kappa \left[\hat{w}(z) \right] \mp \frac{\omega}{\kappa} \mathcal{G}^j_\kappa \left[\hat{\eta} - \hat{\eta}_g \right] 
        \right] \\
    =& \frac{1}{2}  \left[ \frac{1}{\omega \kappa} \frac{\kappa^2 \omega h_\kappa^j}{\kappa} + \frac{1}{\omega \kappa} \frac{\omega \kappa^2 h_\kappa^j}{\kappa}  
        \right] \\
    =&  h_\kappa^j
    \end{align}
\end{subequations}
and thus
\begin{equation}
A_w^i = \frac{\braopket{\wmode}{\op{H}}{\vect{\psi}}}{\braopket{\wmode}{\op{H}}{\wmode}}= \frac{e^{\mp i \omega_\kappa^j t}}{2 \kappa h_\kappa^j}  \left( i \mathcal{G}^j_\kappa \left[\hat{w}(z) \right] \mp \omega_\kappa^j \mathcal{G}^j_\kappa \left[\hat{\eta} - \hat{\eta}_g \right] 
        \right)
\end{equation}
as noted in table \ref{tab:solution-projection}.

%
\section{Proof of orthogonality}\label{secA1}
%

Orthogonality between modes is trivial for modes at different wavenumbers following \eqref{eqn:fourier-orthogonality},
\begin{equation}
\label{eqn:two-mode-general-orthogonality}
    \braopket{\psi^a}{\op{H}}{\psi^b} = \frac{1}{2} \delta_{k_a k_b} \delta_{\ell_a \ell_b} \int_{-D}^0 \hat{u}_a \conj{\hat{u}}_b + \hat{v}_a \conj{\hat{v}}_b + \hat{w}_a \conj{\hat{w}}_b + N^2(z) \hat{\eta}_a \conj{\hat{\eta}}_b \, dz
\end{equation}
and so we need only consider orthogonality of modes at the same wavenumber and thus can drop the $\delta_{k_a k_b} \delta_{\ell_a \ell_b}$ from the proofs that follow.

\subsection{Orthogonality between two geostrophic modes}

Inserting two geostrophic modes from table \ref{tab:solutions} into \eqref{eqn:two-mode-general-orthogonality} leads to the condition that,
\begin{equation}
    \braopket{\gmode^a}{\op{H}}{\gmode^b} =\frac{1}{2} \int_{-D}^0 \frac{g^2 \kappa^2}{f^2} F_g^a F_g^b + N^2 (z) G_g^a G_g^b \, dz = 0
\end{equation}
when $a \neq b$. Orthogonality follows immediately from the orthogonality of the $F_g$ modes \eqref{eqn:f-norm-geostrophic} and the $G_g$ modes \eqref{eqn:g-norm-geostrophic}.


\subsection{Orthogonality between geostrophic and wave modes}

First note that using integration by parts we can show that
\begin{equation}
    \int_{-D}^0 g F_\kappa F_g \, dz = \int_{-D}^0 g h_\kappa \partial_z G_\kappa F_g \, dz = - g h_\kappa \int_{-D}^0 G_\kappa \partial_z F_g \, dz, 
\end{equation}
using that $F_\kappa = h_\kappa \partial_z G_\kappa$ and the boundary conditions $G_\kappa(0) = G_\kappa(-D) = 0$. Inserting one wave mode and one geostrophic mode from table \ref{tab:solutions} into \eqref{eqn:two-mode-general-orthogonality} leads to the condition that
\begin{subequations}
    \begin{align}
        \braopket{{\wmode}^a}{\op{H}}{\gmode^b} =& \pm \frac{k}{\omega_\kappa^a} e^{i k x \pm i \omega_\kappa^a t} \int_{-D}^0 g F_\kappa^a F_g^b - h_\kappa^a N^2 G_\kappa^a G_g^b \, dz \\
        =& \mp \frac{k}{\omega_\kappa^a} h_\kappa^a e^{i k x \pm i \omega_\kappa^a t} \int_{-D}^0  G_\kappa^a \left( g \partial_z F_g^b +  N^2 G_g^b \right)\, dz \\
        =& 0
    \end{align}
\end{subequations}
when $a \neq b$. The last step follows using that $g \partial_z F_g^b = - N^2 G_g^b$.



\subsection{Orthogonality between two wave modes}

Inserting two wave modes from table \ref{tab:solutions} into \eqref{eqn:two-mode-general-orthogonality} leads to the condition that
\begin{subequations}
    \begin{align} \nonumber
        &\braopket{{\wmode}^a}{\op{H}}{\gmode^b} \\ =& \frac{1}{2} e^{i(\omega_\kappa^a- \omega_\kappa^b)t} \int_{-D}^0 \left[ 1+ \frac{f^2}{\omega_\kappa^a \omega_\kappa^b}\right] F_\kappa^a F_\kappa^b + \kappa^2 h_\kappa^a h_\kappa^b G_\kappa^a G_\kappa^b + \frac{\kappa^2 h_\kappa^a h_\kappa^b}{\omega_\kappa^a \omega_\kappa^b} N^2 G_\kappa^a G_\kappa^b \, dz \\ \label{eqn:two-wave-proof}
        =& \frac{1}{2} e^{i(\omega_\kappa^a- \omega_\kappa^b)t} \int_{-D}^0 \left[ 1+ \frac{f^2}{\omega_\kappa^a \omega_\kappa^b}\right] \left( F_\kappa^a F_\kappa^b + \kappa^2 h_\kappa^a h_\kappa^b G_\kappa^a G_\kappa^b \right) dz 
    \end{align}
\end{subequations}
where we used that $\int (N^2-f^2) G_\kappa^a G_\kappa^b dz = 0$ for $a \neq b$ from \eqref{eqn:g-norm-igw}. Using integration by parts and $G_\kappa(0) = G_\kappa(-D) = 0$, this condition also leads to
\begin{equation}
    \int_{-D}^0 F_\kappa^a F_\kappa^b + h_\kappa^a h_\kappa^b \kappa^2 G_\kappa^a G_\kappa^b \, dz = 0
\end{equation}
which means the integral in \eqref{eqn:two-wave-proof} vanishes, concluding the proof.

\subsection{Orthogonality between the mda and inertial modes}

Table \ref{tab:solutions} shows that mda modes have only non-trivial isopycnal deviation, while inertial oscillations have only non-trivial horizontal velocity, and thus \eqref{eqn:two-mode-general-orthogonality} is trivially satisfied.

\end{appendices}

\bibliographystyle{unsrtnat}

\begin{thebibliography}{29}
\providecommand{\natexlab}[1]{#1}
\providecommand{\url}[1]{\texttt{#1}}
\expandafter\ifx\csname urlstyle\endcsname\relax
  \providecommand{\doi}[1]{doi: #1}\else
  \providecommand{\doi}{doi: \begingroup \urlstyle{rm}\Url}\fi

\bibitem[Vallis(2006)]{vallis2006-book}
Geoffrey~K. Vallis.
\newblock \emph{{Atmospheric and Oceanic Fluid Dynamics: Fundamentals and
  Large-Scale Circulation}}.
\newblock Cambridge University Press, 1 edition, 2006.
\newblock \doi{10.1017/9781107588417}.

\bibitem[Garrett and Munk(1972)]{garrett1972-gfd}
Christopher Garrett and Walter Munk.
\newblock {Space-Time scales of internal waves}.
\newblock \emph{Geophysical Fluid Dynamics}, 3\penalty0 (1):\penalty0 225 --
  264, 1972.
\newblock \doi{10.1080/03091927208236082}.

\bibitem[Kelly(2016)]{kelly2016-jpo}
Samuel~M. Kelly.
\newblock {The Vertical Mode Decomposition of Surface and Internal Tides in the
  Presence of a Free Surface and Arbitrary Topography}.
\newblock \emph{Journal of Physical Oceanography}, 46\penalty0 (12):\penalty0
  3777 -- 3788, 2016.
\newblock \doi{10.1175/jpo-d-16-0131.1}.

\bibitem[Olbers(1986)]{olbers1986-igw}
Dirk Olbers.
\newblock {Internal gravity waves}.
\newblock In \emph{Landolt-Börnstein - Numerical data and functional
  relationships in science and technology}, volume~3a, pages 37--82.
  Springer-Verlag, Berlin, 1986.

\bibitem[Smith and Vanneste(2013)]{smith2013-jpo}
K.~Shafer Smith and Jacques Vanneste.
\newblock {A Surface-Aware Projection Basis for Quasigeostrophic Flow}.
\newblock \emph{Journal of Physical Oceanography}, 43\penalty0 (3):\penalty0
  548 -- 562, 2013.
\newblock \doi{10.1175/jpo-d-12-0107.1}.

\bibitem[Hernandez-Duenas et~al.(2014)Hernandez-Duenas, Smith, and
  Stechmann]{hernandez2014-jfm}
Gerardo Hernandez-Duenas, Leslie~M Smith, and Samuel~N Stechmann.
\newblock {Investigation of Boussinesq dynamics using intermediate models based
  on wave–vortical interactions}.
\newblock \emph{Journal of Fluid Mechanics}, 747:\penalty0 247--287, 2014.
\newblock ISSN 0022-1120.
\newblock \doi{10.1017/jfm.2014.138}.

\bibitem[Eden et~al.(2019)Eden, Pollmann, and Olbers]{eden2019-jpo}
Carsten Eden, Friederike Pollmann, and Dirk Olbers.
\newblock {Numerical evaluation of energy transfers in internal gravity wave
  spectra of the ocean}.
\newblock \emph{Journal of Physical Oceanography}, 49\penalty0 (3):\penalty0
  737--749, 2019.
\newblock ISSN 0022-3670.
\newblock \doi{10.1175/jpo-d-18-0075.1}.

\bibitem[Eden et~al.(2020)Eden, Pollmann, and Olbers]{eden2020-jpo}
Carsten Eden, Friederike Pollmann, and Dirk Olbers.
\newblock {Towards a global spectral energy budget for internal gravity waves
  in the ocean}.
\newblock \emph{Journal of Physical Oceanography}, 50\penalty0 (4):\penalty0
  935--944, 2020.
\newblock ISSN 0022-3670.
\newblock \doi{10.1175/jpo-d-19-0022.1}.

\bibitem[Hernández-Dueñas et~al.(2021)Hernández-Dueñas, Lelong, and
  Smith]{hernandez2021-jpo}
Gerardo Hernández-Dueñas, M~Pascale Lelong, and Leslie~M Smith.
\newblock {Impact of Wave-Vortical Interactions on Oceanic Submesoscale Lateral
  Dispersion}.
\newblock \emph{Journal of Physical Oceanography}, 2021.
\newblock ISSN 0022-3670.
\newblock \doi{10.1175/jpo-d-20-0299.1}.

\bibitem[Lelong et~al.(2020)Lelong, Cuypers, and
  Bouruet-Aubertot]{lelong2020-jpo}
M.~Pascale Lelong, Yannis Cuypers, and Pascale Bouruet-Aubertot.
\newblock {Near-Inertial Energy Propagation inside a Mediterranean Anticyclonic
  Eddy}.
\newblock \emph{Journal of Physical Oceanography}, 2020.
\newblock \doi{10.1175/jpo-d-19-0211.1}.

\bibitem[Early et~al.(2021)Early, Lelong, and Sundermeyer]{early2021-jfm}
Jeffrey~J. Early, M.P. Lelong, and M.A. Sundermeyer.
\newblock {A generalized wave-vortex decomposition for rotating Boussinesq
  flows with arbitrary stratification}.
\newblock \emph{Journal of Fluid Mechanics}, 912:\penalty0 A32, 2021.
\newblock ISSN 0022-1120.
\newblock \doi{10.1017/jfm.2020.995}.

\bibitem[Remmel and Smith(2009)]{remmel2009-jfm}
Mark Remmel and Leslie Smith.
\newblock {New intermediate models for rotating shallow water and an
  investigation of the preference for anticyclones}.
\newblock \emph{Journal of Fluid Mechanics}, 635:\penalty0 321--359, 2009.
\newblock ISSN 0022-1120.
\newblock \doi{10.1017/s0022112009007897}.

\bibitem[Chouksey et~al.(2023)Chouksey, Eden, Masur, and
  Oliver]{chouksey2023-jfm}
Manita Chouksey, Carsten Eden, Gökce~Tuba Masur, and Marcel Oliver.
\newblock {A comparison of methods to balance geophysical flows}.
\newblock \emph{Journal of Fluid Mechanics}, 971:\penalty0 A2, 2023.
\newblock ISSN 0022-1120.
\newblock \doi{10.1017/jfm.2023.602}.

\bibitem[Vasylkevych and Žagar(2021)]{vasy2021-qjrms}
Sergiy Vasylkevych and Nedjeljka Žagar.
\newblock {A high‐accuracy global prognostic model for the simulation of
  Rossby and gravity wave dynamics}.
\newblock \emph{Quarterly Journal of the Royal Meteorological Society},
  147\penalty0 (736):\penalty0 1989--2007, 2021.
\newblock ISSN 0035-9009.
\newblock \doi{10.1002/qj.4006}.

\bibitem[O’Neill et~al.(2024)O’Neill, Chelton, Rodríguez, Samelson, and
  Wineteer]{oneill2024-jtech}
Larry~W O’Neill, Dudley~B Chelton, Ernesto Rodríguez, Roger Samelson, and
  Alexander Wineteer.
\newblock {Feasibility of estimating sea surface height anomalies from surface
  ocean currents and winds}.
\newblock \emph{Journal of Atmospheric and Oceanic Technology}, 2024.
\newblock ISSN 0739-0572.
\newblock \doi{10.1175/jtech-d-23-0096.1}.

\bibitem[Wagner and Young(2015)]{wagner2015-jfm}
G.~L. Wagner and W.~R. Young.
\newblock {Available potential vorticity and wave-averaged quasi-geostrophic
  flow}.
\newblock \emph{Journal of Fluid Mechanics}, 785:\penalty0 401--424, 2015.
\newblock ISSN 0022-1120.
\newblock \doi{10.1017/jfm.2015.626}.

\bibitem[Müller(1995)]{muller1995-rg}
Peter Müller.
\newblock {Ertel's potential vorticity theorem in physical oceanography}.
\newblock \emph{Reviews of Geophysics}, 33\penalty0 (1):\penalty0 67--97, 1995.
\newblock ISSN 8755-1209.
\newblock \doi{10.1029/94rg03215}.

\bibitem[Early et~al.(2022)Early, Hernández-Dueñas, Smith, and
  Lelong]{early2022-arxiv}
Jeffrey~J Early, Gerardo Hernández-Dueñas, Leslie~M Smith, and M~Pascale
  Lelong.
\newblock {Exact expressions for available potential energy and available
  potential vorticity}.
\newblock \emph{arXiv}, 2022.
\newblock \doi{10.48550/arxiv.2212.07405}.

\bibitem[Holliday and Mcintyre(1981)]{holliday1981-jfm}
Dennis Holliday and Michael~E. Mcintyre.
\newblock {On potential energy density in an incompressible, stratified fluid}.
\newblock \emph{Journal of Fluid Mechanics}, 107\penalty0 (-1):\penalty0
  221--225, 1981.
\newblock ISSN 1469-7645.
\newblock \doi{10.1017/s0022112081001742}.

\bibitem[Winters and Barkan(2013)]{winter2013-jfm}
Kraig~B. Winters and Roy Barkan.
\newblock {Available potential energy density for Boussinesq fluid flow}.
\newblock \emph{Journal of Fluid Mechanics}, 714:\penalty0 476--488, 2013.
\newblock ISSN 0022-1120.
\newblock \doi{10.1017/jfm.2012.493}.

\bibitem[Schneider et~al.(2003)Schneider, Held, and Garner]{schneider2003-jas}
Tapio Schneider, Isaac~M. Held, and Stephen~T. Garner.
\newblock {Boundary Effects in Potential Vorticity Dynamics}.
\newblock \emph{Journal of the Atmospheric Sciences}, 60\penalty0 (8):\penalty0
  1024--1040, 2003.
\newblock ISSN 0022-4928.
\newblock \doi{10.1175/1520-0469(2003)60<1024:beipvd>2.0.co;2}.

\bibitem[Roullet and Klein(2008)]{roullet2008-jfm}
Guillaume Roullet and Patrice Klein.
\newblock {Available potential energy diagnosis in a direct numerical
  simulation of rotating stratified turbulence}.
\newblock \emph{Journal of Fluid Mechanics}, 624:\penalty0 45--55, 2008.
\newblock ISSN 0022-1120.
\newblock \doi{10.1017/s0022112008004473}.

\bibitem[Munk(1981)]{munk1981-book}
Walter Munk.
\newblock {Internal Waves and Small-Scale Processes}.
\newblock Evolution of Physical Oceanography, pages 264 -- 291. 1981.

\bibitem[Roullet and Klein(2010)]{roullet2010-prl}
Guillaume Roullet and Patrice Klein.
\newblock {Cyclone-Anticyclone Asymmetry in Geophysical Turbulence}.
\newblock \emph{Physical Review Letters}, 104\penalty0 (21):\penalty0 218501,
  2010.
\newblock ISSN 0031-9007.
\newblock \doi{10.1103/physrevlett.104.218501}.

\bibitem[Chelton et~al.(2011)Chelton, Schlax, and Samelson]{chelton2011-pio}
Dudley~B. Chelton, Michael~G. Schlax, and Roger~M. Samelson.
\newblock {Global observations of nonlinear mesoscale eddies}.
\newblock \emph{Progress In Oceanography}, 91\penalty0 (2):\penalty0 167 --
  216, 2011.
\newblock \doi{10.1016/j.pocean.2011.01.002}.
\newblock URL
  \url{http://www.sciencedirect.com/science/article/pii/S0079661111000036}.

\bibitem[Anderson and Killworth(1979)]{anderson1979-dsr}
David L.~T. Anderson and Peter~D. Killworth.
\newblock {Nonlinear propagation of long Rossby waves}.
\newblock \emph{Deep Sea Research Part I: Oceanographic Research Papers},
  26\penalty0 (9):\penalty0 1033 -- 1049, 1979.
\newblock \doi{10.1016/0198-0149(79)90046-3}.
\newblock URL
  \url{http://www.sciencedirect.com/science/article/pii/0198014979900463}.

\bibitem[Charney and Flierl(1981)]{charney1981-book}
J.G. Charney and G.R. Flierl.
\newblock {Oceanic analogues of large-scale atmospheric motions}.
\newblock Evolution of Physical Oceanography, pages 504 -- 548. 1981.

\bibitem[Petviashvili(1980)]{petviashvili1980-jetp}
V.I. Petviashvili.
\newblock {Red spot of Jupiter and the drift soliton in a plasma}.
\newblock \emph{Journal of Experimental and Theoretical Physics Letters},
  32:\penalty0 619 -- 622, 1980.

\bibitem[Hoskins et~al.(1985)Hoskins, McIntyre, and
  Robertson]{hoskins1985-qjrms}
B.~J. Hoskins, M.~E. McIntyre, and A.~W. Robertson.
\newblock {On the use and significance of isentropic potential vorticity maps}.
\newblock \emph{Quarterly Journal of the Royal Meteorological Society},
  111\penalty0 (470):\penalty0 877--946, 1985.
\newblock ISSN 0035-9009.
\newblock \doi{10.1002/qj.49711147002}.

\end{thebibliography}

\end{document}